\documentclass[a4paper,english]{article}
\usepackage[utf8x]{inputenc}
\usepackage{textcomp}
\usepackage{amsmath}
\usepackage{graphicx}

\makeatletter

\pdfpageheight\paperheight
\pdfpagewidth\paperwidth

\providecommand{\tabularnewline}{\\}

\@ifundefined{date}{}{\date{}}

\usepackage[english]{babel}
\usepackage[colorinlistoftodos]{todonotes}
\usepackage{epstopdf}
\usepackage{authblk}
\usepackage{rotating}
\usepackage[graphicx]{realboxes}
\usepackage{varwidth}

\title{A Unifying Framework for the Identification of Motor Primitives}

\author[1]{Enrico Chiovetto}
\author[2,3]{Andrea s'Avella}
\author[1]{Martin Giese}
\affil[1]{Section for Computational Sensomotorics, Department of Cognitive Neurology, Hertie Institute for Clinical Brain Research, Centre for Integrative Neuroscience, University Clinic Tuebingen, Tuebingen, Germany.}
\affil[2]{Department of Biomedical and Dental Sciences and Morphofunctional Images, University of Messina, Messina, Italy.}
\affil[3]{Laboratory of Neuromotor Physiology, Santa Lucia Foundation, Rome, Italy.}

\usepackage{babel}

\usepackage{babel}

\makeatother

\usepackage{babel}
\begin{document}
\maketitle 
\begin{abstract}
A long-standing hypothesis in neuroscience is that the central nervous
system accomplishes complex motor behaviors through the combination
of a small number of motor primitives. Many studies in the last couples
of decades have identified motor primitives at the kinematic, kinetic,
and electromyographic level, thus supporting modularity at different
levels of organization in the motor system. However, these studies
relied on heterogeneous definitions of motor primitives and on different
algorithms for their identification. Standard unsupervised learning
algorithms such as principal component analysis, independent component
analysis, and non-negative matrix factorization, or more advanced
techniques involving the estimation of temporal delays of the relevant
mixture components have been applied. This plurality of algorithms
has made difficult to compare and interpret results obtained across
different studies. Moreover, how the different definitions of motor
primitives relate to each other has never been examined systematically.
Here we propose a comprehensive framework for the definition of different
types of motor primitives and a single algorithm for their identification.
By embedding smoothness priors and specific constraints in the underlying
generative model, the algorithm can identify many different types
of motor primitives. We assessed the identification performance of
the algorithm both on simulated data sets, for which the properties
of the primitives and of the corresponding combination parameters
were known, and on experimental electromyographic and kinematic data
sets, collected from human subjects accomplishing goal-oriented and
rhythmic motor tasks. The identification accuracy of the new algorithm
was typically equal or better than the accuracy of other unsupervised
learning algorithms used previously for the identification of the
same types of primitives. 
\end{abstract}

\section*{Introduction}

A fundamental challenge in neuroscience is to understand how the central
nervous system (CNS) controls the large number of degrees-of-freedom
(DOF) of the musculoskeletal apparatus to perform a wide repertoire
of motor tasks and behaviors. A long-standing hypothesis is that the
CNS relies on a modular architecture in order to simplify motor control
and motor learning {[}1-3{]}. Many studies in recent years have indeed
shown that kinematic {[}4-5{]}, kinetic {[}6-7{]} and electromyographic
(EMG) patterns {[}8-11{]} underlying complex movements can be approximated
by the combinations of a small number of components, usually referred
to as motor primitives or motor synergies. The identification of such
components has typically been carried out by applying unsupervised
learning algorithms, including principal component analysis (PCA),
independent component analysis (ICA) {[}5, 12-15{]}, non-negative
matrix factorization (NMF) {[}15, 16{]} or other methods inspired
by such algorithms {[}17{]}. While these classical methods are based
on instantaneous mixture models, that linearly combine a set of basis
vectors time-point by time-point, more advanced techniques have also
been proposed that involve the estimation of temporal delays between
relevant mixture components {[}15, 18-21{]}. This multitude of underlying
mathematical models complicates the comparison of results from different
studies on motor primitives. In addition, even for the same mathematical
models often multiple algorithms for the estimation of motor primitives
have been proposed, and it is not always clear if their results are
comparable. This further complicates the comparison of the results.
Finally, how the different definitions of motor primitives relate
to each other has never been systematically examined. We propose in
this article a new comprehensive framework for the definition of motor
primitives and a new algorithm for their identification. We show that
many different definitions of spatial, temporal and spatiotemporal
primitives given in the literature can be derived from a single generative
model that is known as “anechoic mixture” and relies on the combination
of components that can be shifted in time. When the delays of all
primitives are constrained to be zero, the anechoic model reduces
to the instantaneous linear combination model, which underlies the
definition of spatial or temporal synergies, usually identified by
PCA, ICA or NMF. Similarly, when specific equality and non-negativity
constraints are imposed on its parameters, the model can describe
spatiotemporal synergies {[}9, 15{]}. In addition to this unification
of models, we present a new identification algorithm that estimates
motor primitives, according to the different definitions, with an
accuracy that is equal or even better than the standard techniques
that are commonly used for the identifications of these motor primitives.
The robustness of this new algorithm results from an integration of
smoothness priors and appropriate constraints in the underlying generative
model. The new algorithm has been validated by assessing its identification
performance both on simulated data sets, for which the properties
of the primitives and of the corresponding combination parameters
were known, and on experimental EMG and kinematic data sets, collected
from human participants accomplishing goal-oriented and rhythmic motor
tasks. The new algorithm is publically available, is provided as a
toolbox in MATLAB (The Mathworks, Natick, MA) and can be downloaded
for free from www.compsens.uni-tuebingen.de. In this way, we aim to
provide the field of motor control with a new usable and robust tool
for the identification of motor primitives, helping to reduce the
inconsistencies and incompatibilities between the different generative
models.

\section*{Methods }

\subsubsection*{Generative models for the description of motor primitives }

We give in this section a brief survey of the definitions of motor
primitives and of the corresponding generative models that have been
used in the literature for the investigation of the modular organization
of motor behavior. The different approaches can be subdivided into
different groups, according to the model features that are assumed
to be invariant across conditions. In the following, a matrix $\textbf{X}^{l}$
indicates the data corresponding to a specific trial $l$ ($0\leq l\leq L$),
where $L$ is the total number of trials collected during an experiment.
Each row of $\textbf{X}^{l}$ represents a specific degree of freedom
(DOF) of the system under investigation (for instance an angular trajectory
associated with a specific joint in the case of kinematic data, or
the electrical signals associated with the contraction of a specific
muscle in the case of EMG data). Each column of $\textbf{X}^{l}$
contains the values assumed by the different DOF at a particular point
in time. Unless the size of the matrix is explicitly mentioned, from
now on $\textbf{X}^{l}$ will be assumed to have $M$ rows (number
of DOF) and $T$ columns (equivalent to the number of time samples
in one trial). Signals are supposed to be sampled at constant sampling
frequency and to have duration $T_{s}$. In the following, an individual
column that corresponds to the time point $t$ of $\textbf{X}^{l}$
will also be signified by the column vector $\textbf{x}^{l}(t)$,
so that $\textbf{X}^{l}=[\textbf{x}^{l}(1),...,\textbf{x}^{l}(T)]$.
The components of these vectors will be indicated by the variables
$x_{n}^{l}(t)$. In the following, we give an overview of different
models for motor primitives that have been proposed previously in
the literature.

\subsubsection*{Spatial primitives}

One classical definition of motor primitive is based on the idea that
groups of DOF might show instantaneous covariations, reflecting a
coordinated recruitment of multiple muscles or joints. This implies
the assumption that the ratios of the signals characterizing the different
DOF remain constant over time. This type of movement primitive has
been applied in particular in muscle space, where muscle synergies
have been defined as weighted groups of muscle activations {[}3, 10,
22{]}. Such synergies have also been referred to as “synchronous”
synergies, since the different muscles are assumed to be activated
synchronously without muscle-specific time delays. Consistent with
this definition is the following generative model that, from now on,
will be referred to as ‘spatial decomposition’:

\begin{equation}
\textbf{x}^{l}(t)=\sum_{p=1}^{P}\textbf{w}_{p}\cdot c_{p}^{l}(t)+residuals
\end{equation}

In this equation the vectors $\textbf{x}^{l}(t)$ indicate the values
of the individual DOF at time point $t$ (assuming discrete time steps,
$1\le t\le T$) in trial number $l$. The column vectors $\textbf{w}_{p}$
define the ‘spatial patterns’ of the muscle synergies, which are assumed
to be invariant over trials. The number of primitives is $P$, and
the scalars $c_{p}^{l}(t)$ are the time-dependent mixing weights
of the primitives. The mixing weights, as well as the residuals, are
different in every trial. Processed EMG data typically consists of
time series of non-negative signals, i.e. $x_{m}^{p}(t)\geq0$, for
$1\le t\le T$ and $1\le m\le M$. In these models it is typically
also assumed that the components of the mixture model (1) (except
for the residuals) are non-negative, i.e. $c_{p}^{l}(t)\geq0$ and
$w_{p,m}\geq0$ (where the subscript $m$ indicates the $m$-th element
of the vector $\textbf{w}_{p}$).

\subsubsection*{Temporal primitives}

An alternative way to characterize motor primitives is based on the
idea that they express invariance across time, defined by basic temporal
patterns or functions $s_{p}(t)$ that are combined or superposed
in order to reconstruct a set of temporal signals. Temporal components
based on this definition have been identified in kinematic {[}4-5,
18{]}, dynamic {[}6{]} and EMG {[}8, 11-12{]} space. The underlying
generative model (which from now on we will refer to as ‘temporal
decomposition’) is mathematically described as:

\begin{equation}
x_{m}^{l}(t)=\sum_{p=1}^{P}c_{mp}^{l}\cdot s_{p}(t)+residuals
\end{equation}

In this equation $x_{m}^{l}(t)$ is the value of the $m$-th DOF at time
$t$ in trial number $l$, and the corresponding scalar mixing weights
$c_{mp}^{l}$ change between trials of different types (experimental
conditions). The temporal primitives $s_{p}(t)$, however, are assumed
to be invariant over trials. $P$ signifies the total number of temporal
primitives. Another more elaborated model of this type has been proposed
in {[}19, 21-23{]}. This model allows for temporal shifts between
the temporal basis functions for different DOF. This can be interpreted
as reflecting, for example, delays between the activation of different
muscles within the same primitive. Mathematically, this model is characterized
by the equations:

\begin{equation}
x_{m}^{l}(t)=\sum_{p=1}^{P}c_{mp}^{l}\cdot s_{p}(t-\tau_{mp}^{l})+residuals
\end{equation}

The time shifts between the basis functions for the different degrees
of freedom are captured by the variables $\tau_{mp}^{l}$. The time
delays and linear mixing weights are typically assumed to vary over
trials, while it is assumed that the basis functions $s_{p}(t)$ are
invariant, as in model (2). Like for model (1), inequality constraints
can be imposed on the mixing weights in models of type (2) and (3),
for example to account for the non-negativity of EMG signals.

\subsubsection*{Spatiotemporal (time-varying) primitives}

Spatiotemporal (or time-varying) primitives have been proposed as
a way to model EMG components that are invariant in both, space and
time {[}9, 15, 24{]}. Moreover, for each primitive additional temporal
delays are admitted, similar to model (3). This results in the following
generative model (referred to as ‘spatiotemporal decomposition’),
where $\textbf{x}^{l}(t)$ signifies again the time-dependent column
vector of the DOF as function of time:

\begin{equation}
\textbf{x}^{l}(t)=\sum_{p=1}^{P}c_{p}^{l}\cdot\textbf{w}_{p}(t-\tau_{p}^{l})+residuals
\end{equation}

Again, the mixing weights $c_{p}^{l}$ and the delays $\tau_{p}^{l}$
change between different trial types while the functions $\textbf{w}_{p}(t)$
are assumed to be invariant, defining the primitives or muscle synergies.
The time-varying synergies and the corresponding mixing weights have
typically been assumed to be non-negative {[}15{]}, although also
models with unconstrained parameters have been applied to model phasic
EMG activity {[}9{]}.

\subsubsection*{Space-by-time primitives }

Recently, Delis and colleagues {[}25{]} proposed a new synergy model
for EMG data, which they named ‘space-by-time decomposition’. This
model merges the definitions of spatial and temporal components into
a new definition of primitives that is given by the following equation:

\begin{equation}
\textbf{x}^{l}(t)=\sum_{p=1}^{P_{tp}}\sum_{q=1}^{P_{sp}}s(t-\tau_{pq}^{l})\cdot c_{pq}^{l}\cdot\textbf{w}_{q}+residuals
\end{equation}

In this model, $\textbf{w}_{q}$ and $s_{p}(t)$ define the trial-independent
spatial and temporal components as in models (1) and (2), while the
mixing weights $c_{pq}^{l}$ and time delays $\tau_{pq}^{l}$ are
trial-dependent. The constants $P_{tp}$ and $P_{sp}$ indicate the
numbers of temporal and spatial components. Since the model was originally
designed to account for EMG data, Delis and colleagues assumed all
parameters of the model equation (5) to be non-negative (except for
the time delays).

\subsubsection*{Unifying model}

All previously discussed models can be derived as special instantiations
of a single model, called ‘anechoic mixture model’. This type of model
is known from acoustics, where it is applied for modeling of acoustic
mixtures in reverberation-free rooms {[}26-29{]}. This model assumes
typically a set of $R$ recorded acoustic signals $y_{r}(t)$ that
are created by the superposition of $U$ acoustic source functions
$f_{u}(t)$, where time-shifted versions of these source functions
are linearly superposed with the mixing weights $a_{ru}$. The time
shifts are given by the time delays $\tau_{ru}$. This models the
fact that for a reverberation-free room the signals from the acoustic
sources arrive receiver with different time delays and attenuated
amplitudes, which are dependent on the distances between the acoustic
sources and the receivers. The corresponding generative model has
the following form (for $1\leq r\leq R$ ):

\begin{equation}
y_{r}(t)=\sum_{u=1}^{U}a_{ru}\cdot f_{u}(t-\tau_{ru})+residuals
\end{equation}

\subsubsection*{Equivalence between the unifying model and the other models }

By addition of appropriate constraints, the anechoic mixture model
(6) can be made equivalent to all previously discussed models for
motor primitives. This becomes obvious by the following considerations:

a) Identifying the signals of type $y_{r}(t)$ with the components
of the vectors $\mathbf{x}^{l}(t)$, i.e. $y_{r}(t)=x_{m(r)}^{l(r)}(t)$
(where the integer functions $l(r)$ and $m(r)$ define a one-to-one
mapping between the $m$-th degree of freedom in trial $l$ and the
corresponding signal $y_{r}(t)$ (with $1\leq r\leq M\cdot L$),
and constraining the time delays $\tau_{ru}$ to be zero, one obtains
the model (1). Since in this model the weight vectors $\mathbf{w}_{p}$
are assumed to be invariant over trials, all mixing weights $a_{rp}$
belonging to the same DOF and primitive number $P$ have to be equal
and independent of the trial number, so that $a_{rp}=w_{p,m(r)}$,
where the function $m(r)$ returns the number of the DOF that belongs
to index $r$ independent of the trial number. The time-dependent
mixing coefficients $c_{p}^{l}(t)$ of the model (1) correspond to
the source functions $f_{u}$ of the model (6), thus $f_{u}(t)=c_{p(u)}^{l(u)}(t)$
where here the index $u$ runs over all combinations of the indices
$p$ and $l$, thus $1\leq u\leq U=P\cdot L$ and where the integer
functions $l(u)$ and $p(u)$ establish mappings between the number
of the source function in model (6) and the time-dependent mixing
weights in model (1). Non-negativity constraints can be added for
the model parameters $a_{rp}$and the functions $f_{u}(t)$, e.g.
for the modeling of EMG data.

b) If one identifies the source functions in model (6) with the temporal
primitive functions $s_{p}(t)$, i.e. $f_{p}(t)=s_{p}(t)$, $1\leq p\leq P$
and again constrains the delays $\tau_{ru}$ to be zero, equation
(6) becomes equivalent to model (2). In this case, the mixing weights
$a_{rp}$are equated with the mixing coefficients $c_{mp}^{l}$ in
model (2), where the index $r$ runs over all combinations of $m$
and $l$, formally $a_{rp}=c_{m(r),p}^{l(r)}$, with appropriately
chosen integer functions $m(r)$ and $l(r)$. Like for model (1),
the components of the data vector have to be remapped over DOF and
trials according to the relationship $y_{r}(t)=x_{m(r)}^{l(r)}$(t).
Again, non-negativity constraints can be added for the parameters
$a_{rp}$and to the source functions $f$.

c) Dropping the constraints $\tau_{ru}=0$ in the equivalences described
in b), and equating the delays in model (3) according to the relationship
$\tau_{rp}=\tau_{m(r),p}^{l(r)}$, makes model (6) equivalent to model
(3).

d) Introducing individual sets of basis functions for the different
DOF, grouping them into vectors and equating the mixing weights and
temporal delays for the components of each vector, transforms model
(6) into the model (4). On the level of the time-dependent basis functions,
this equivalence can be mathematically described by the equation $f_{u}(t)=w_{p(u),m(u)}(t)$,
where $w_{p,m}$ corresponds to the component of the basis function
vector $\mathbf{w}_{p}(t)$ that belongs to the $m$-th DOF, and where
the integer functions $m(u)$ and $p(u)$ establish a one-to-one mapping
between the indices of the basis functions in the two models and the
number of the associated DOF. This assignment is independent of the
trial index $l$. The index $r$ in (6) runs over all combinations
of DOF and trial numbers, thus $1\leq r\leq M\cdot L$. The integer
functions $m(r)$ and $l(r)$ assign the corresponding trial number
and DOF to the index $r$ in the model (6). Thus, the assignment equation
for the data vector is again given by $y_{r}(t)=x_{m(r)}^{l(r)}(t)$
for the $m$-th DOF in the $l$-th trial. The requirement that all mixing
weights and temporal delays belonging to the same basis function vector
$\mathbf{w}_{p}$ are equal is equivalent to a set of equality constraints,
which can be captured by the equation systems $a_{ru}=c_{p(r)}^{l(r)}$
and $\tau_{ru}=\tau_{p(r)}^{l(r)}$ . Again, non-negativity constraints
can be added, if necessary.

e) In order to establish equivalence with the model (5), the data
vectors of the models are mapped onto each other according to the
relationship $y_{r}(t)=x_{m(r)}^{l(r)}(t)$, where again $l(r)$ and
$m(r)$ are integer mapping functions that assign the $r$-th element
of the data vector of the model (6) to the $m$-th DOF of the data vector
$\mathbf{x}^{l}$ for the $l$-th trial in (5) with $1\leq r\leq M\cdot L$.
Model (5) has a total of $P_{sp}\cdot P_{tp}$ temporal basis functions,
where however the functional forms of the basis functions for different
indices $q$ (i.e. different spatial components) for the same $p$
(i.e. same temporal component) just differ by time shifts. This is
equivalent to an equality constraint for these functions, which can
mathematically be characterized in the form $f_{u}(t)=s_{p(u)}(t)$,
with $1\leq u\leq P_{tp}$ and the index functions $p(u)$ and $q(u)$
that map the index $u$ in the model (6) onto the indices of the temporal
and spatial primitive in (5). Since all indices with the same $p(u)$
are mapped onto the same basis function sp the last equation specifies
an equality constraint. With the same integer mapping functions, finally,
also the relationship between the mixing weights can be established,
which is given by the equation $a_{ru}=c_{p(u),q(u)}^{l(r)}\cdot w_{q(r),m(u)}$,
 where $w_{q,m}$ is the $m$-th element for the vector $\mathbf{w}_{q}$.
The last equation specifies a bilinear constraint for the weight parameters
of the model (6). Using the same notation, the equivalence between
the delays is established by the equation system $\tau_{ru}=\tau_{p(u),q(u)}^{l(r)}$.
A summary of the established equivalences between the general model
(6) and the other models is given in Table 1.

\begin{table}
\centering \rotatebox{90}{\begin{varwidth}{\textheight} \scalebox{0.91}{
\begin{tabular}{|c|c|c|c|c|}
\cline{2-5} 
\multicolumn{1}{c|}{} & Spatial (1)  & Temporal (2) or (3)  & Spatiotemporal (4)  & Space-by-time (5)\tabularnewline
\multicolumn{1}{c|}{} &  &  &  & \tabularnewline
\multicolumn{1}{c|}{} & $\textbf{x}^{l}(t)=\sum\limits_{p=1}^{P}\textbf{w}_{p}\cdot c_{p}^{l}(t)$  & $x_{m}^{l}(t)=\sum\limits_{p=1}^{P}c_{mp}^{l}\cdot s_{p}(t-\tau_{mp}^{l})$  & $\textbf{x}^{l}(t)=\sum\limits_{p=1}^{P}c_{p}^{l}\cdot\textbf{w}_{p}(t-\tau_{p}^{l})$  & $\textbf{x}^{l}(t)=\sum\limits_{p=1}^{P_{tp}}\sum\limits_{q=1}^{P_{sp}}s(t-\tau_{pq}^{l})\cdot c_{pq}^{l}\cdot\textbf{w}_{q}$\tabularnewline
\hline 
Anechoic (6)  &  &  &  & \tabularnewline
 & $y_{r}(t)=x_{m(r)}^{l(r)}(t)$  & $y_{r}(t)=x_{m(r)}^{l(r)}(t)$  & $y_{r}(t)=x_{m(r)}^{l(r)}(t)$  & $y_{r}(t)=x_{m(r)}^{l(r)}(t)$\tabularnewline
 &  &  &  & \tabularnewline
$y_{r}(t)=\sum\limits_{u=1}^{U}a_{ru}\cdot f_{u}(t-\tau_{ru})$  & $f_{u}(t)=c_{p(u)}^{l(u)}(t)$  & $f_{p}(t)=s_{p}(t)$,  & $f_{u}(t)=w_{p(u),m(u)}(t)$  & $f_{u}(t)=s_{p(u)}(t)$,\tabularnewline
 &  &  &  & \tabularnewline
 & $a_{rp}=w_{p,m(r)}$  & $a_{rp}=c_{m(r),p}^{l(r)}$  & $a_{ru}=c_{p(r)}^{l(r)}$  & $a_{ru}=c_{p(u),q(u)}^{l(r)}\cdot w_{q(r),m(u)}$ \tabularnewline
 &  &  &  & \tabularnewline
 & $\tau_{ru}=0$  & $\tau_{ru}=0$ or $\tau_{rp}=\tau_{m(r),p}^{l(r)}$  & $\tau_{ru}=\tau_{p(r)}^{l(r)}$  & $\tau_{ru}=\tau_{p(u),q(u)}^{l(r)}$\tabularnewline
\hline 
\end{tabular}} \caption{Constraints that make the primitive models (1), (2), (3), (4) and
(5) equivalent to the general anechoic model (6). See text for details.}
\end{varwidth}} 
\end{table}

\subsection*{An efficient algorithm for the identification of motor primitives
within the unified framework }

The solution of anechoic demixing problems is a well-known topic in
unsupervised learning, with close relationship to methods such as
ICA and blind source separation (cf. e.g. {[}30-31{]}). Numerous algorithms
have been proposed to solve this problem for the most general case
where the functions $f$ are assumed to be elements of relatively
general function spaces. For the under-determined case (in which the
number of signals/sensors is smaller than the number of sources) well-known
algorithms include information maximization approaches {[}32{]} and
frequency, or time-frequency methods {[}33-34{]}, such as the DUET
algorithm {[}29{]}. Other work for the under-determined case is summarized
in Ogrady et al. {[}35{]}, Arberet et al. {[}36{]} and Cho and Kuo
{[}37{]}. The over-determined case (where the signals outnumber the
sources) is much more interesting for dimensionality reduction applications,
but has been addressed more rarely. Harshman and colleagues {[}38{]}
developed an alternating least squares (ALS) algorithm for this problem
(Shifted Factor Analysis). Their method was later revised and improved
by Mørup and colleagues {[}39{]}, who exploited the Fourier shift
theorem and information maximization in the complex domain (SICA,
Shifted Independent Component Analysis). More recently, Omlor and
Giese {[}19{]} developed a framework for blind source separation,
starting from stochastic time-frequency analysis that exploited the
marginal properties of the Wigner-Ville spectrum. The discussed algorithms
solve the anechoic demixing problem for the most general case, at
the cost that they are computationally expensive. All algorithms for
blind source separation require the identification of a large number
of parameters. Given model (6), $T\cdot U$ parameters need to be
identified to represent all sources $f_{u}(t)$, and for each trial
$l$, $M\cdot U$ weights $a_{ru}$ and $M\cdot U$ delays $\tau_{ru}$.
Given a whole data set, this results in a total number of parameters
to be identified that is $(T+2M\cdot L)\cdot U$, where typically
$T\gg M,U,L$ (with $T,M,U$ and $L$ indicating the total numbers
of time samples, DOF, sources, and trials). For applications
in motor control, the relevant signals are subject to additional constraints,
which can be exploited for the derivation of more efficient algorithms.
Signals in motor control are typically smooth. This allows to reduce
considerably the complexity of the anechoic demixing problem and to
devise algorithms that are more robust than those developed for general
purposes. We present in this section a unifying algorithm for standard
anechoic demixing, which can be used for the identification of the
parameters associated with the unconstrained model (6). The general
version of this algorithm, which from now on we will refer to as FADA
(Fourier-based Anechoic Demixing Algorithm), was introduced in a previous
study to identify primitives defined according to eq. (3) {[}18{]}.
Here we describe how this algorithm can be extended by inclusion of
additional constraints that make it suitable for the identification
of the parameters associated with different models for primitives.
The time-courses of signals related to body movements (trajectories
as well as EMG traces) often are relatively smooth and thus can be
approximated well by anechoic mixtures of smooth signals {[}18{]}.
This smoothness of the source functions $f(u)$ can be expressed by
appropriate priors that help to stabilize the source separation problem.
Smooth temporal sources can be approximated by truncated Fourier expansions.
Consequently, each source can be approximated by $K$ complex Fourier
coefficients, where$K$ is typically far below the Nyquist limit ($K\ll T/2$).
Consequently, the number of parameters to identify drops remarkably
to $(K+2M\cdot L)\cdot U$. This decreases substantially the computational
costs of the parameter estimation and make it more robust. When the
temporal signals $y_{r}(t)$ and sources $f_{u}(t)$ are assumed to
be band-limited they can be approximated by truncated Fourier expansions
of the form:

\begin{equation}
y_{r}(t)=\sum_{k=-K}^{K}c_{rk}e^{\frac{2\pi ikt}{T_{s}}}
\end{equation}

and

\begin{equation}
f_{u}(t-\tau_{ru})\cong\sum_{k=-K}^{K}\nu_{uk}e^{-ik\tau_{ru}}e^{\frac{2\pi ikt}{T_{s}}}
\end{equation}

where $c_{rk}$ and $\nu_{uk}$ are complex constants ($c_{rk}=\left|c_{rk}\right|e^{i\varphi_{c_{rk}}}$
and $\nu_{uk}=\left|\nu_{uk}\right|e^{i\varphi_{\nu_{uk}}}$ ), and
where $i$ is the imaginary unit. The positive integer $K$ is determined
by Shannon’s theorem according to the limit frequency of the signals,
and $T_{s}$ is the temporal duration of the signal. The source separation
algorithm tries to ensure that the source functions $f_{u}(t)$ are
uncorrelated over the distributions of the approximated signals. This
implies $E\left\{ f_{u}(t)\cdot f_{u'}(t')\right\} =0$ for $u\neq u'$
and any pair $t\neq t'$. For the corresponding Fourier coefficients
this implies $E\left\{ \nu_{uk}\cdot\nu_{u'k'}\right\} =0$ for $u\neq u'$
and any pair $k\neq k'$ . Combining equation (6), (7) and (8) we
obtain by comparison of the terms for the same frequency

\begin{equation}
c_{rk}=\sum_{u=1}^{U}a_{ru}\cdot\nu_{uk}e^{-ik\tau_{ru}}
\end{equation}

From this follows with $E\left\{ \nu_{uk}\cdot\nu_{u'k'}^{*}\right\} =E\left\{ \left|\nu_{uk}\right|^{2}\right\} \cdot\delta_{uu'}$
the equation:

\begin{equation}
\begin{split}\left|c_{rk}\right|^{2} & =E\left\{ \left|c_{rk}\right|\right\} \\
 & =\sum_{u=1}^{U}\sum_{u'=1}^{U}a_{ru}a_{ru'}E\left\{ \nu_{uk}\cdot\nu_{u'k'}^{*}\right\} e^{-ik(\tau_{ru}-\tau_{ru'})}\\
 & =\sum_{u=1}^{U}a_{ru}^{2}E\left\{ \left|\nu_{uk}\right|^{2}\right\} \\
 & =\sum_{u=1}^{U}\left|a_{ru}\right|^{2}\left|\nu_{uk}\right|^{2}
\end{split}
\end{equation}

The symbol $*$ indicates the conjugate of a complex number. The derivation
of this equation replaces the expectations of the Fourier coefficients
$c_{rk}$ with their deterministic values and treats the source weights
$a_{rk}$ as deterministic trial-specific variables. This can be justified
if these mixture weights are estimated separately from the sources
in an EM-like procedure. Empirically, however, we obtain reasonable
estimates of the model components based on equation (10) also using
other methods (see below). Since the signals $f_{u}(t)$ and $y_{r}(t)$
are real the corresponding Fourier coefficients fulfil $c_{rk}=c_{r,-k}^{*}$
and $\nu_{uk}=\nu_{u,-k}^{*}$. Thus the demixing problem needs to
be solved only for parameters with $k\geq0$.

The previous considerations motivate the following iterative algorithm
for the identification of the unknown parameters in model (6). After
random initialization of the estimated parameters, the following steps
are carried out iteratively until convergence: 
\begin{enumerate}
\item Compute the absolute values of the coefficients $c_{rk}$ and solve
the following equations: 
\begin{equation}
\left|c_{rk}\right|^{2}=\sum\limits_{u=1}^{U}\left|a_{ru}\right|^{2}\left|\nu_{uk}\right|^{2}
\end{equation}
with $r=0,1,\ldots R$ and $k=0,1,\ldots K$. In our study we exploited
non-negative ICA {[}40{]} to solve this equation. In the distributed
version of the software equation (10) can also be solved via non-negative
matrix factorization {[}34, 41{]}. 
\item Initialize for all pairs and iterate the following steps: 

\begin{enumerate}
\item Update the phases of the Fourier coefficients of the sources, defined
as $\varphi_{\nu_{uk}}=angle(\nu_{uk})=arctan(Im(\nu_{uk})/Re(\nu_{uk}))$
by solving the following non-linear least square problem

\begin{equation}
\underset{\mathbf{\Phi}}{\mathrm{min}}\left\Vert \mathbf{C}-\mathbf{Z}(\mathbf{\mathbf{\Phi}})\right\Vert _{F}^{2}
\end{equation}
where $(\mathbf{C})_{rk}=c_{rk}$, $(\mathbf{Z})_{rk}=\sum\limits_{u=1}^{U}a_{ru}e^{-ik\tau_{uk}}\left|\nu_{uk}\right|e^{i\varphi_{\nu_{uk}}}$
and indicates the Frobenius norm. (In order to avoid cluttered notation,
for the function $\mathbf{Z}(.)$ only the arguments with relevance
for the optimization are explicitly written). 

\item Keeping the identified source functions $f_{u}(t)$ constant, identify
for each signal $y_{r}(t)$ the weights $a_{ru}$ and delays $\tau_{ru}$
by minimization of the following cost functions: 
\begin{equation}
\underset{\mathbf{\mathrm{\mathbf{a}}}_{r},\mathbf{\boldsymbol{\tau}}_{r}}{\arg\min}\left\Vert y_{r}(t)-\mathbf{f}(t,\mathbf{\mathbf{\mathbf{\boldsymbol{\tau}}}}_{r})'\mathrm{\mathbf{a}}_{r}\right\Vert _{F}^{2}
\end{equation}

\end{enumerate}
\end{enumerate}
Optimization with respect to $\mathbf{a}_{r}$ and $\boldsymbol{\tau}_{r}$
is feasible, assuming uncorrelatedness of the functions $f_{u}$ and
independence of the time delays {[}42{]}. The column vector $\mathbf{a}_{r}$
concatenates all weights associated with DOF $r$, i.e $\mathbf{a}_{r}=\left[a_{r1},\ldots,a_{rU}\right]'$.
The vector function $\mathbf{f}_{r}(t,\boldsymbol{\tau}_{r})=\left[f_{1}(t-\tau_{r1}),\ldots,f_{U}(t-\tau_{rU})\right]$
concatenates the functions $f_{u}$, shifted by the time delays associated
with the $r$-th DOF.

The original version of the FADA algorithm was designed to solve the
source separation problems without constraints. Additional constraints,
such as the non-negativity of the parameters or additional equality
constraints for the weights and delays can be easily added, due to
the modular structure of the algorithm. The following sections briefly
describe the additional constraints that were introduced in order
to implement the identification of the parameters of models (1), (2),
(3), (4) and (5).

\subsubsection*{Non-negativity of the primitives}

For the case where the primitives $f_{u}$ can assume only non-negative
values, equation (10) cannot be derived in the way discussed above,
and the expression of the non-negativity constraints in the Fourier
representation is not straightforward. We decided, instead to estimate
the time-dependent values of $f_{u}(t)$ directly, taking the inequality
constraint $f_{u}(t)\geq0$ for discretely sampled values of into
account. This results in the following algorithm: Starting from random
values of the parameters, the following three steps are iterated until
convergence: 
\begin{enumerate}
\item Update of the absolute values of the Fourier coefficients $\left|\nu\right|_{uk}$ of
the primitives $f_{u}$, assuming their phases $\varphi_{\nu_{uk}}$and
the mixing weights $a_{ru}$ are known, by solving the non-linear
constrained optimization problem:

\begin{equation}
\begin{aligned} & \underset{\mathbf{N}}{\mathrm{minimize}} &  & \left\Vert \mathbf{C}-\mathbf{Z}(\mathbf{N})\right\Vert _{F}^{2}\\
 & \text{subject to} &  & f_{u}(\mathbf{N},t)\geq0,\;u=1,2,\ldots U\;\mathrm{and}\;t=1,\ldots T.
\end{aligned}
\end{equation}

In order to avoid cluttered notation, for the functions $\mathbf{Z}(.)$
and $f_{u}(.)$ only the arguments with relevance for the optimization
are explicitly written.) The matrix is defined as in (12) and $(\mathbf{Z})_{rk}=\sum\limits_{u=1}^{U}a_{ru}e^{-ik\tau_{uk}}\left|\nu_{uk}\right|e^{i\varphi_{\nu_{uk}}}$,
with $(\mathbf{N})_{uk}=\left|\nu_{uk}\right|$. 

\item Assuming the other parameters are fixed, update the phases $\varphi_{\nu_{uk}}$
of the Fourier coefficients of the primitives by solving the non-linear
constrained optimization problem 
\begin{equation}
\begin{aligned} & \underset{\mathbf{\Phi}}{\mathrm{minimize}} &  & \left\Vert \mathbf{C}-\mathbf{Z}(\mathbf{\Phi})\right\Vert _{F}^{2}\\
 & \text{subject to} &  & f_{u}(\mathbf{\Phi},t)\geq0,\;u=1,2,\ldots U\;\mathrm{and}\;t=1,\ldots T.
\end{aligned}
\end{equation}

Remind that the Fourier coefficients $\nu_{u0}$ are real so that it
is sufficient to regard consider only $k=1,\ldots,K$. 

\item Update weights and delays as in the unconstrained version of FADA
by solving the optimization problem (13). 
\end{enumerate}

\subsubsection*{Non-negativity of the mixing coefficients}

Non-negativity of the scaling coefficients $a_{ru}$ of the primitives
can be easily imposed in the algorithm. In (13) the scaling coefficients
are determined, assuming that primitives and temporal delays are known,
solving a least squares problem. The same optimization problem can
be solved adding the linear inequality constrains $a_{ru}\geq0$,
$\forall r,u$, resulting in a non-negative least squares problem
for the weights $a_{ru}$.

\subsubsection*{Identification of spatial synergies}

The FADA algorithm presented above can be used to identify not only
temporal, but also spatial primitives. This can be achieved simply
by transposing the data matrix $\mathbf{X}$ and constraining all
the delays in the algorithm to be equal to 0. In this way indeed,
the FADA algorithm identifies a set of invariant spatial (instead
of temporal) vectors, interpreting the elements of each vector $\mathbf{x}(t)$
as a series of time points. Although there is no theoretical evidence
for the existence of any smoothness relation between the values of
the different DOF at a given time instant $t$ (so that the smoothness
assumptions of FADA on the data are satisfied), it will be shown in
the next sections how the algorithm can however still provide identification
performance at least as good as those associated with other standard
machine learning techniques.

\subsubsection*{Identification of spatiotemporal synergies }

For the identification of spatiotemporal synergies, constraints for
the parameters have to be set according to model (4). In a first step,
for each DOF $m$ in the data set $P$ source functions $f_{p}$ are
assigned, resulting in a total of $M\cdot P$ independent source functions.
The following three steps are then carried out iteratively until convergence: 
\begin{enumerate}
\item The optimal delays $\tau_{p}^{l}$ for each spatiotemporal primitive
are found, for each trial $l$, applying a matching pursuit procedure
{[}43-44{]}, consisting of an iterative search for a set of time-shifted
primitives that best match the data. For each primitive, the scalar
product between the original data and the time-shifted primitive is
computed, testing all possible time delays between 0 and $T$-1. The
primitive and delay associated with the highest scalar product is
then selected and its contribution is subtracted from the data. Then
the same procedure is repeated for the remaining primitives on the
residual of the data. This search is repeated until all delays have
been determined. 
\item The combination coefficients $c_{p}^{l}$ are updated by minimizing,
for each trial $l$, the difference between the original data and
the reconstruction, estimated exploiting model (4) and assuming that
the source functions $f_{u}$ and the delays $\tau_{ru}$ are known. 
\item Assuming that the weights and the delays are known from the previous
steps, the functions $f_{u}$, which correspond to the components
of the spatiotemporal primitives $w_{p}(t)$ are updated. The Fourier
coefficients of the corresponding source function are determined in
the same fashion as for the original FADA algorithm without constrains.
Non-negativity constraints for the primitives and weights can be imposed
in the same way as described above. 
\end{enumerate}

\subsubsection*{Identification of space-by-time synergies }

We developed a new algorithm for the identification of space-by-time
primitives, exploiting the core of FADA algorithm (the mapping onto
the Fourier space) for the identification of the temporal primitives
associated with the space-by-time factorization. Similar to Delis
and colleagues {[}25{]}, this algorithm was also designed for the
processing of EMG-like data and all the parameters in model (5) (with
the exception of the delays) are constrained to be non-negative. Given
the data matrix $\mathbf{X}$, in the first step of the algorithm
$P_{sp}$ spatial primitives $\mathbf{w}_{q}$ are identified, applying
non-negative matrix factorization {[}41{]}. Then the FADA algorithm
is applied to $\mathbf{X}$ in order to identify $P_{tp}$ non-negative
temporal primitives $s_{p}(t)$. In the second step of the algorithm,
the spatial primitives are kept constant, while temporal primitives,
weights and delays are updated. The algorithm consists of the iteration
of the two following steps: 
\begin{enumerate}
\item The Fourier coefficients of the functions $s_{p}(t)$ are updated
as in the constrained FADA algorithm, by minimizing the difference
between the Fourier coefficients of the original data and the linear
combination of the corresponding Fourier coefficients. 
\item Weights and delays are updated minimizing the difference between the
original data and the estimates provided by model (5). The optimal
delays $\tau_{qp}^{l}$ are found for each trial $l$, following a
matching pursuit procedure. Similarly, the weights $c_{qp}^{l}$ are
identified, solving for each trial a constrained linear least-squares
problem. 
\end{enumerate}
To minimize the risk of finding local minima, we always ran the FADA
algorithm 10 times on the same data set with different random initial
conditions and we considered only the solutions that provided the
lowest error in the reconstruction of the original data. To test whether
these solutions actually represented points close to the global minimum,
we computed the average similarity between the sets of primitives
identified at the end of each run of the algorithm (see below for
the definition of similarity). Indeed, a high level of similarity
between these solutions can be considered as a strong sign that, with
very high probability, these solutions are close to the optimal one.
In the case, for instance, of an artificial mixture of non-negative
temporal components based on model (2), we found that the average
similarity between the identified primitives was very high (0.98 on
a scale where 1 indicates perfect matching (see equation 16). This
high level of similarity allows to rule out the hypothesis that the
solutions provided by FADA represent local minima. For the identification
of temporal, spatiotemporal or space-by-time primitives, the number
of harmonics $K$ was always set according to the following procedure:
We computed the average spectrum from all signals within the data
and defined $K$ as the closest integer that approximates the product
of the signal duration $T_{s}$ and the average band-width $B$ of
the data set. This number was alwayssmaller thanthe limit $K_{max}$
imposed by the the Nyquist-Shannon theorem. Differently, in the case
of spatial primitives we always set $K=K_{max}$.

\subsubsection*{Other identification algorithms}

The FADA algorithm was benchmarked against other unsupervised learning
methods for the extraction of synergies. For data based on the synchronous
unconstrained generative models (1) and (2) we used the fastICA algorithm
{[}45-46{]} (function ‘fastica.m’ of the corresponding toolbox). We
examined the performance of fastICA after reducing the dimensionality
of the data using principal component analysis. For the fastICA algorithm
we found the level of similarity between original and identified synergies
depended on the number of principal components and it reached the
highest value when the number of principal components was equal to
the number of synergies in the data. Based on this observation we
always set the number of principal components to the number of identified
synergies. Non-negative matrix factorization {[}34, 41{]} (NMF) was
used to identify the model parameters for synchronous mixture with
non-negative components and mixing weights. We used the Matlab function
“nnmf.m”, implementing the matrix multiplication update rule version
of the algorithm introduced by Lee and Seung {[}34, 41{]}. For data
relying on model (3) we used the anechoic demixing algorithm (AnDem)
developed by Omlor and Giese {[}19{]} and the shifted ICA algorithm
(SICA) by Morup {[}39{]} (downloaded from http://www2.imm.dtu.dk/pubdb/views/publication\_details.php?id=5206
{[}47{]}). For anechoic demixing with non-negativity constraints we
used an anechoic NMF algorithm (ANMF) developed by Omlor and Giese
{[}19{]} and the shifted NMF (sNMF) by Morup and colleages {[}48{]},
who kindly provided us with the Matlab implementation of the algorithm.
To extract time-varying synergies we used the modified NMF algorithm
developed by d’Avella and colleagues {[}9, 15{]} (stNMF, standing
for spatiotemporal non-negative matrix factorization). Finally, we
compared the performance of the FADA algorithm for the identification
of temporal and spatial primitives from the space-by-time model with
the performance of the sample-based non-negative matrix tri-factorization
algorithm (sNM3F) developed by Delis and colleagues {[}25{]}.

\subsubsection*{Generation of the simulated data }

For the quantitative assessment of algorithm’s performance we simulated
kinematic and EMG data sets that were compatible with equations (1),
(2), (3), (4) and (5). Each of these data sets approximated coarsely
the properties of real biological signals. Each data set consisted
of $M$-dimensional trajectories with $T$ time steps and $L$repeated
trials. Synthesized EMG signals were constrained to be non-negative,
like real EMG signals after rectification and filtering. All generative
models were based on a set of statistically independent temporal waveforms.
These waveforms (source functions, or synergies) corresponded to the
time-dependent combination coefficients $c_{p}^{l}(t)$ in model (1),
to the temporal signals $s_{p}(t)$ in models (2) and (3) and (5),
and to the components of the vector function in model in (4). For
the generation of the unconstrained sources we drew 100 random samples
from a normal distribution (Matlab function “randn.m”) and low-pass
filtered with a Butterworth filter with normalized cut-off frequency
equal to 0.15 (Matlab functions “butter” and “filtfilt”). This procedure
allowed to generate band-limited, smooth sources mimicking the typical
properties of real kinematic or kinetic trajectories with a length
of $T$ = 100 time samples. For the generation of EMG-like sources
we produced spike trains from a multi-dimensional stochastic renewal
process {[}49{]}, and convolved them with a Gaussian function. The
renewal process was a homogeneous Poisson process characterized by
random inter-spike intervals drawn from an exponential distribution
with mean 1/$\lambda$, where the rate parameter of the Poisson process
was given by $\lambda$ = 40 Hz. Based on the random inter-spike intervals,
spike trains with length $T$ = 100 were generated. Each spike train
was then convolved with a Gaussian filter kernel with a standard deviation
of 8 discrete time steps. The generated source signals were used to
construct the synergies in the generative models (2), (3), (4) and
(5). The weight vectors $\mathbf{w}$ in (1) and (5) were obtained
by drawing $M$ random samples from a uniform distribution over the
interval {[}-40 40{]} for the unconstrained case, and from an exponential
distribution with mean 20 for the cases with non-negativity constraints.
Examples of generated primitives are shown in Fig 1. For kinematic
(unconstrained) data sets based on model (2) and (3) the values of
the coefficients $c_{mp}$ were drawn from a uniform distribution
over the interval {[}-20, 20{]}. For EMG-like data sets based on the
models (2), (3), (4) and (5) the scaling coefficients were drawn from
exponential distributions with mean 10. For all the models with time
delays $\tau\neq0$, the delays were drawn from exponential distributions
with mean 20 and rounded to the nearest integer. The time delays sampled
from this distribution with values larger than $T$ = 100 were taken
modulo to map them back to the interval {[}0, $T$-1{]}. Noisy data
was derived by adding signal-dependent noise {[}50-53{]} to the generated
data. The noise was drawn from a Gaussian distribution with mean 0
and standard deviation $\sigma=\alpha\left|x(t)\right|$, where $\alpha$
is a scalar and $x(t)$ is the value of the noiseless data at the
time instant $t$. The slope $\alpha$ was computed though an iterative
procedure. Starting from $\alpha=0$, its value was iteratively increased
of a predefined increment until the level of the difference $1-R^{2}$
(where the parameter$R^{2}$ describes the level of similarity between
two data sets, see below) reached a predefined value. For each noiseless
data set, three data sets were generated with$1-R^{2}$ levels equal
to 0.05, 0.15, 0.25 and 0.35). For each generative model, 20 noiseless
data sets were simulated that were consistent with equations (1) to
(5), randomly selecting synergies, scaling coefficients and time delays.
The number of synergies $P$ was always set to 4 and the number of
simulated DOFs was 10. The number of simulated trials $L$ was 25.
The time duration of each trial was assumed to be $T_{s}=1$ and the
sampling frequency was set to 100 Hz.

\subsubsection*{Experimental kinematic and EMG data}

We assessed the identification performance of each algorithm also
on actual experimental kinematic and EMG data. The kinematic data
set consisted of flexion angle trajectories of the body joints recorded
from human actors walking with different emotional styles (neutral,
happy and sad). These data were used in previous work on emotional
gaits {[}20, 23, 54{]}. From this data set, unconstrained temporal
primitives were identified with the FADA and the anechoic demixing
algorithm. EMG data consisted of previously published recordings {[}9{]}
obtained from 16 arm muscles during arm reaching movements. These
muscle activation patterns were used to investigate the production
of behaviors through combination of muscle synergies. The recorded
EMG raw signals were digitally full-wave rectified, low-pass filtered
(20 Hz cut-off) and integrated within time bins of 10 ms. All EMGs
signals in the data set were resampled to fit a 75-point time window
(0.75s).

\subsubsection*{Assessment of algorithm performance }

For each algorithm we assessed three different performance measures,
quantifying the capability of each algorithm to identify the original
movement primitives, the original activation coefficients and the
original delays in comparison to the parameters used to generate the
data. The similarity between original and extracted primitives was
quantified by computing the maximum of the scalar products between
original and identified primitives, taking the maximum over all possible
time delays in cases where model contained temporal shifts. Let $p_{1}(t)$
and $p_{2}(t)$ signify the compared primitives or source functions
(discretely sampled in time) and that these signals are normalized
so that their norm is one. The similarity measure is defined by the
scalar product of these normalized signals, where one of them is time-shifted
with time delay , where this delay is optimized by maximizing the
similarity measure. For models without time shifts the time delays
are constrained to be 0. Mathematically the correlation measure is
given by the equation:

\begin{equation}
S=\underset{\tau}{\max}\underset{\tau}{\sum}p_{1}(t)\cdot p_{2}(t-\tau)
\end{equation}

For the case of time-varying synergies (model 4) the compared signals
were vector-valued. In this case the scalar product of the vectors
in (16) was taken for each (delayed) time step and the signals were
normalized ensuring that for $\underset{t}{\sum}\left|\mathbf{p}_{j}((t)\right|^{2}=1$,
for $j=1,2$. The similarity measure takes values between -1 and 1,
where the value 1 corresponds to the situation that both source function
have identical shape (except for maybe a time delay). In order to
establish correspondence between the individual primitives of the
generative model and the identified primitives, we first computed
the similarity measure $S$ for all possible pairings of the primitives
and chose then the pairing with the highest similarity score. For
this purpose, first the pairing with the highest similarity score
was determined and removed from the original and reconstructed model.
Then this procedure was repeated for the second-best matching pair
of the remaining set of primitives, and so forth. This procedure was
iterated until all primitives had been matched. The similarity between
original and identified coefficients (or time delays) was assessed
by computing the correlation coefficients between activation coefficients
(temporal delays) of the matched primitives. These correlation coefficients
were then averaged across all matched pairs of primitives. We also
defined a measure of similarity between original and reconstructed
data sets. Since the generated and experimental kinematic or EMG patterns
and the residuals of the reconstruction of the patterns by synergy
combinations were multivariate time-series, a measure of the goodness
of the reconstruction (typically a ratio of two variances) had to
be defined. We used the “total variation” {[}55{]}, defined as the
trace of the covariance matrix of the signals. A multivariate measure
$R^{2}$ for the explained data variance is then given by the expression

\begin{equation}
R^{2}=1-\frac{\sum\limits_{l=1}^{L}\left\Vert \mathbf{X}^{l}-\mathbf{X}_{rec}^{l}\right\Vert }{\sum\limits_{l=1}^{L}\left\Vert \mathbf{X}^{l}-\bar{\mathbf{X}^{l}}\right\Vert }
\end{equation}

where each $\mathbf{X}^{l}$was the matrix of the actual data associated
with trial $l$, $\mathbf{X}_{rec}^{l}$ the reconstructed values
by the fitted model, and where $\bar{\mathbf{X}}$ is the matrix of
the mean values of the data over trial $l$. The statistical distributions
of all similarity measures described above for randomized data were
assessed for each algorithm and each data set. In order to calculate
baseline levels for these similarity measures ($\Sigma_{b}$), we
first randomly generated 20 independent sets of synergies, coefficients,
and delays (where appropriate) using the corresponding generative
model. The similarities between the identified synergies (activation
coefficients or delays) between these randomly generated sets were
then computed. The obtained similarities were then averaged over all
20 simulations, resulting in a baseline value $\Sigma_{b}$ for the
corresponding similarity measure. The similarity measures $\Sigma$
resulting from the comparison between the identified and the simulated
primitives were than transformed into a normalized similarity measure
according to the formula:

\begin{equation}
\Sigma_{norm}=\frac{\Sigma-\Sigma_{b}}{1-\Sigma_{b}}
\end{equation}

The normalized similarity measure $\Sigma_{norm}$ takes the value
one for perfect similarity, and it is zero if the similarity matches
the average similarity between two randomly generated data sets.

\subsubsection*{Statistical analysis }

All tested measures were normally distributed according to a Chi-square
goodness-of-fit test. Student’s t-test was used to test whether the
reconstructions accuracies and the levels of similarities were statistically
different from chance level. Differences between more than two groups
were statistically tested by two-way ANOVAs (with Algorithm and Noise
Level as factors), where appropriate. Post-hoc analysis was conducted
with Tukey-Kramer test, when necessary and appropriate. As level of
significance for the rejection of the null hypotheses in this study
we chose 5

\section*{Results }

\subsection*{Comparisons of algorithm performance on simulated data sets }

To assess algorithm performance we simulated ground-truth data sets
based on the mixture models described in equations (1), (2), (3) ,(4)
and (5). The aim of our comparison was to show that the FADA algorithm
can identify mixture parameters at least as well as other well-known
unsupervised learning methods. Fig 2A shows the average performance
(\textpm SD) of the FADA and the fastICA algorithm applied to mixtures
of unconstrained synchronous synergies, similar to the ones illustrated
in Fig 1A. The bar plots indicate the reconstruction accuracy measure
$R^{2}$ and the normalized similarity measures for the extracted
synergies $S_{norm}$, averaged across 20 data sets for five different
levels of signal–dependent noise. Asterisks indicate significant differences
according to post-hoc testing between average values, obtained with
different algorithms for the same level of noise. The figure shows
that both algorithms provide a good level of reconstruction accuracy
and resulted in an accurate estimation of the original model parameters.
Normalized accuracy measures were typically larger than 0.5, and the
similarity measures for the recovered primitives and weighting coefficients
were always significantly larger than chance level (t(19)$>$9.93, p$<$0.001).
The two-factor ANOVAs revealed a significant main effect for the factor
Noise Level for both the reconstruction accuracy and the identification
of the primitives (F(4,190) ≥ 5.08, p$<$0.001). We found a significant
main effect for the factor Algorithm for all tested parameters (F(1,190)≥11.92,
p$<$0.001). The interaction between the two factors was significant
for and the similarity of the primitives (F(4,190)≥5.08, p$<$0.001).
The post-hoc analysis revealed that, for the same level of noise,
the fastICA and FADA algorithm did not provide significantly different
identification performance, neither for the identification of the
primitives nor for the weighting coefficients (p$>$0.05), although fastICA
provided always significantly higher reconstruction accuracy (p$<$0.05).

\begin{figure}[!ht]
\begin{centering}
\includegraphics[width=0.85\columnwidth]{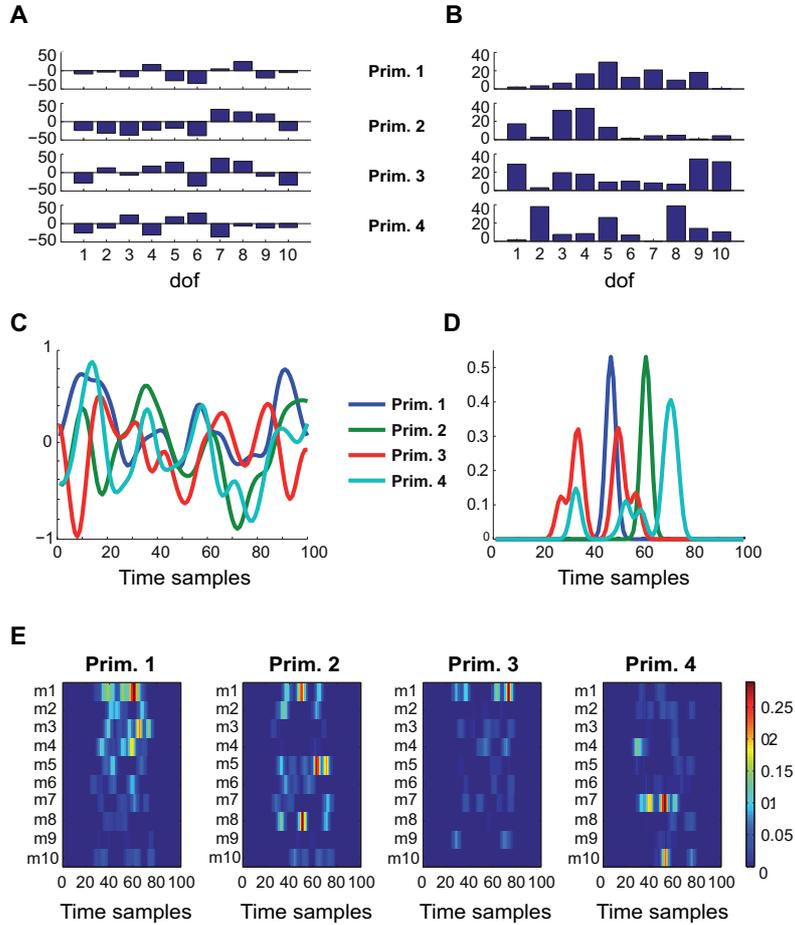} 
\par\end{centering}

\caption{Examples of artificial primitives used for the generation of the ground-truth
data sets. (A) Unconstrained spatial primitives, associated with model
(1). (B) Non-negative spatial primitives, associated with model (1)
and (5). (C) Unconstrained temporal components used in models (2)
and (3). (D) Non-negative (EMG-like) temporal components associated
with model (5). (E) Time-varying non-negative primitives associated
with model (4). }
\label{fig:Fig1} 
\end{figure}

Similarly to Fig 2A, Fig 2B depicts the identification performance
of the algorithms applied to mixtures based on model (1), but synthesized
with non-negative parameters. In this case, we compared the FADA algorithm
to non-negative matrix factorization (NMF), as the fastICA does not
provide a way to constrain parameters to be non-negative. Even in
this case both algorithms provided a good fit of the data and very
accurate estimates of the original primitives and mixture weights.
Not surprisingly, performance of both algorithms degraded with increasing
noise, more remarkably than in Fig 2A. ANOVAs indicated a significant
main effect of the factor Algorithm on the similarity of the weighting
coefficients (F(1,190)=23.14, p$<$0.001). Also the main effect of the
factor Noise Level was significant for both, $R^{2}$ and levels of
normalized similarities (F(4,190)≥20.85, p$<$0.001). The interaction
of both factors was significant only for (F(4,190)=5.51,p$<$0.001).
The post-hoc analysis showed that only in one case (25\% of noise)
NMF performed better than the FADA algorithm in terms of the identification
of the weight coefficients (p=0.74). Differentl;y, NMF provided significantly
lower reconstruction accuracy (p=0.01) for the highest level of noise
(35\%). All average values in Fig 2B were significantly above chance
level (t(19)≥6.85, p$<$0.001).

\begin{figure}[!ht]
\includegraphics[width=0.9\columnwidth]{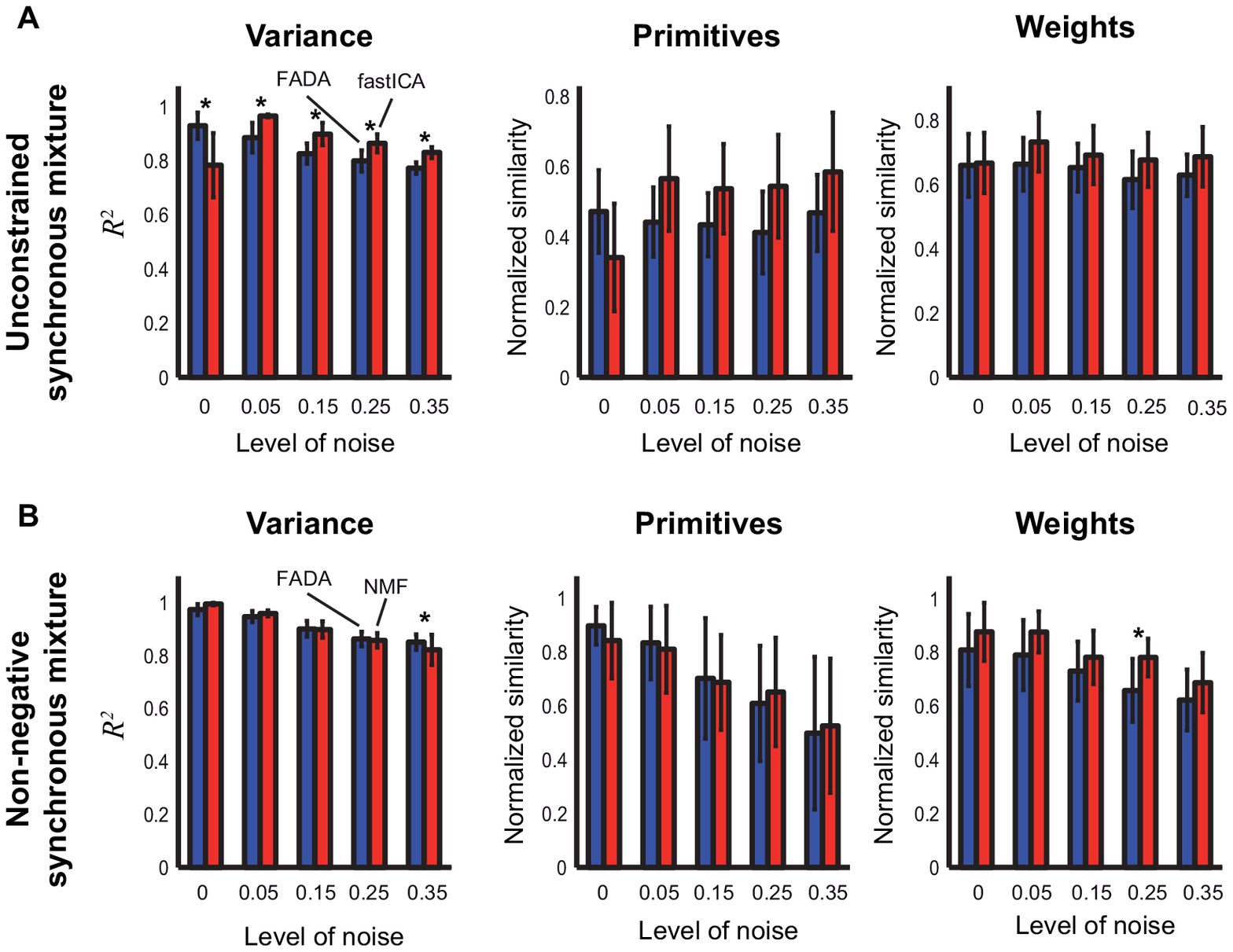} \caption{Spatial primitives. Identification performance (mean \textpm{} SD)
of the Fourier-based anechoic demixing algorithm (FADA), fast independent
component analysis (factICA), and non-negative matrix factorization
(NMF) applied to artificial data sets, which are corrupted by different
amounts of noise, where data was simulated by combining spatial primitives
according to the generative model (1). The number of DOFs in each
spatial primitive was set to $M$ = 10. (A) Level of variance explained
with the extracted parameters, and similarities between original and
identified primitives, and for the corresponding combination coefficients
for unconstrained data (see Methods for details). (B) Level of variance
accounted for with the extracted parameters, similarities between
original and identified spatial primitives, and corresponding combination
coefficients for non-negative data. }
\label{fig:Fig2} 
\end{figure}

Fig 3 shows the identification performance of the FADA, fastICA and
NMF algorithms applied to synchronous mixtures based on model (2).
Fig 3A shows qualitatively that, for the case of unconstrained mixtures,
the level of reconstruction accuracy and the level of similarity of
the primitives were modulated by the level of noise. In contrast,
noise seems to have no significant effect on the estimation of the
weighting coefficients. ANOVAs confirmed a significant main effect
for the factor Noise level for the accuracy of reconstruction and
the similarity of the estimated primitives with the original for the
reconstruction accuracy of the data and the estimation of the weights
(F(1,190)≥9.67, p$<$0.001). The interaction between Algorithm and Noise
Level was significant only for $R^{2}$ (F(4,190)=5.12, p$<$0.001).
Post-hoc testing revealed that the FADA algorithm performed significantly
worse than fastICA in approximating the noisiest data sets (p$<$0.001)
and in terms of the identification of the weighting coefficients (p=0.004)
for one tested noise level (15\%). For all other cases the identification
performance of the FADA and fastICA algorithm did not significantly
differ (p$>$0.05). Fig 3B shows the results of the comparison between
the FADA algorithm and NMF applied to non-negative data. The differences
in performance between the two methods for the same noise levels were
very small. Correspondingly, ANOVAs showed that the factor Algorithm
had a significant main effect only for the reconstruction accuracy
$R^{2}$ (F(1,190)=25,99,p$<$0.001), while the factor Noise had significant
main effects for all three tested measures (F(4,190)≥17.38 p$<$0.001).
Post-hoc testing revealed that the FADA algorithm approximated the
original data with significantly higher reconstruction accuracy than
fastICA, only for the data were corrupted with the two highest levels
of noise (p$<$0.05). Taken together, Figs 2 and 3 show that, when applied
to data based on synchronous models, FADA was in general able to provide
identification performance comparable to those provided by the fastICA
and NMF algorithms for the model (2). In terms of identification of
the actual parameters, Tthe FADA algorithm had worse performance than
the fastICA algorithm only in two single cases concerning the identification
of the weighsfor large noise only for the case of mixtures based on
the combination of unconstrained spatial primitives. Interestingly,
the variability associated with the similarities between original
and identified parameters is higher in Fig. 3A than in Fig 3B. This
is most probably due to an increase of regularization in the algorithms
introduced by the non-negativity constraints imposed on the model
parameters. For the lowest levels of noise (25\% and 35\%) NMF provided
significantly higher reconstruction accuracy (p$<$0.05). All measures
in Fig 3 were significantly above chance level (t(19)≥8.86, p$<$0.001).

\begin{figure}[!ht]
\includegraphics[width=0.9\columnwidth]{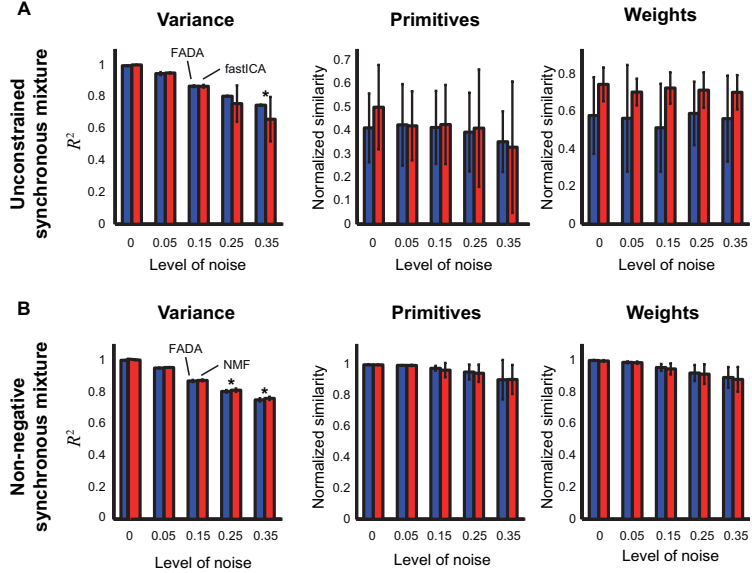} \caption{Instantaneous mixtures (without delays). Identification performance
of the FADA, factICA, and NMF algorithm applied to artificial data
sets generated as instantaneous mixtures of temporal synergies, defined
by the generative model (2). (A) Level of variance explained by the
extracted parameters, source similarities between original and identified
primitives, and similarities of corresponding combination coefficients
for unconstrained data. (B) Same plot for non-negative data.}
\label{fig:Fig3} 
\end{figure}

Fig 4 shows a comparison between the identification performance of
the FADA algorithm and the alternative methods AnDem, SICA, AnNMF,
and sNMF for data generated with model (3) assuming unconstrained
(Fig 4A) and non-negative anechoic (Fig 4B) mixtures. In addition
to the similarity measures assessed before, we also quantified the
similarity between original and identified delays. For the unconstraint
case (Fig 4A), FADA performed qualitatively better than both AnDem
and SICA. For all levels of noise it provided a higher level of data
approximation quality and higher normalized similarities for primitives
and weighting coefficients. FADA and AnDem provided comparable performance
in the identification of the delays. When comparing FADA and AndDem,
ANOVAs revealed a significant main effect of the factor Algorithm
on $R^{2}$ and on the approximation quality for the primitives and
the weights (F(4,190)≥210.5, p$<$0.001). The noise level affected only
$R^{2}$ and the similarity of the primitives (F(4,190)≥12.85, p$\ll$0.001).
The interaction between noise and the type of algorithm was significant
for the estimation of the primitives (F(4,190)≥5.85, p$\ll$0.001). Post-hoc
analysis revealed that FADA provided significantly higher $R^{2}$
values, as well as higher similarities of primitives and weighting
coefficients (p$<$0.05). Compared to the SICA algorithm, the FADA algorithm
showed higher approximation quality of the original data, for almost
all levels of noise, and higher similarities between original and
identified primitives, weights and delays. ANOVAs confirmed a significant
a significant main effect of the factor Algorithm for the estimation
of all parameters (F(4,190)≥10.57, p$<$0.01), while the influence of
the factor Noise Level was sikgnificant only for $R^{2}$ and the
similarities of the weights (F(4,190)≥3.85, p$<$0.01). Post-hoc testing
revealed that the FADA algorithm provided significantly higher $R^{2}$
values than the SICA algorithm (p$<$0.001). For the three highest level
of noise the FADA algorithm also the estimation of the time delays
was more accurate (p$<$0.001). All measures in Fig 4A were significantly
above chance level (t(19)≥3.23, p$<$0.001).

Fig 4B shows the results for the performances of the FADA, AnNMF and
sNMF algorithms for data that are derived from non-negative anechoic
mixtures. Even in this case the FADA algorithm qualitatively provided,
for all levels of noise, higher values of $R^{2}$ and similarity
between original and identified primitives in comparison to the AnNMF
algorithm. ANOVAs revealed a significant main effect of the factor
Algorithm for $R^{2}$ and the similarities of primitives and delays
with the generative model parameters (F(4,190)≥6.64, p$<$0.05). The
ANOVAs revealed also a significant main effect of the factor Noise
Level for $R^{2}$ and all other estimated model parameters (F(4,190)≥2.69,
p$<$0.05). The interaction between the Noise level Algorithm was significant
only for $R^{2}$ and the estimation accuracy of the delays (F(4,190)≥4.29,
p$<$0.01). The post-hoc testing revealed that the FADA algorithm provided
significantly higher $R^{2}$ values than the AnNMF algorithm (p$<$0.001),
higher similarity of the primitives for noisy data sets (p$<$0.05) as
well as higher similarities between original and identified delays
(p$<$0.05) for three level of noise (0\%, 5\% and 15\%). Comparing the
FADA algorithm and sNMF, ANOVAs resulted in a significant main effect
of the factor Algorithm for all measures, except for the similarity
between original and identified delays (F(4,190)≥10.34,p$<$0.01). The
main effect of the factor Noise Level was significant for all parameters
(F(4,190)≥4.41,p$<$0.01). The interaction between both factors was significant
for $R^{2}$ and the identification accuracy of the time delays (F(4,190)≥2.96,p$<$0.05).
Post-hoc testing showed that the FADA algorithm always resulted in
higher reconstruction accuracy and more accurate estimates of the
primitives than sNMF (p$<$0.05), with the only exception of one level
of noise (15\%). All similarity measures in Fig 4B were significantly
above chance level (t(19)≥30.8, p$<$0.01), except for the reconstruction
accuracy provided by sNMF for the most noisy data sets (t(19)=0.25,
p=0.80).

\begin{figure}[!h]
\begin{centering}
\includegraphics[width=0.65\columnwidth]{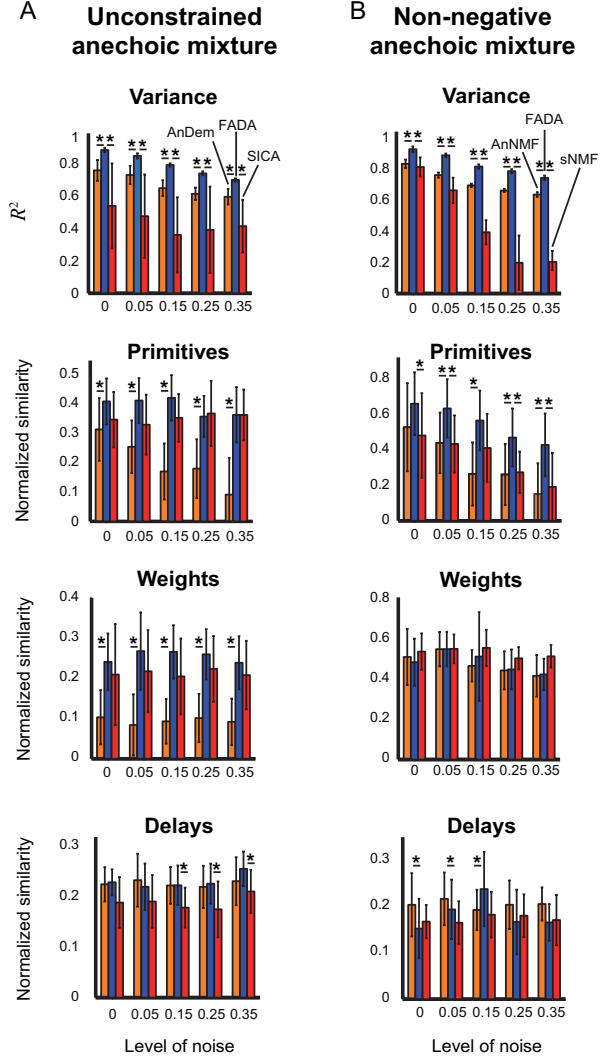} 
\par\end{centering}

\caption{Mixtures with time delays. Identification performance of the FADA
algorithm, the anechoic demixing algorithm by Omlor and Giese (AnDem),
anechoic demixing with non-negativity constraints (AnNMF), shifted
independent component analysis (SICA) and shifted non-negative matrix
factorization (sNMF) algorithm applied to artificial data sets, obtained
by combining temporal synergies linearly with time shifts as described
by model (3). (A) Level of variance that explained with the extracted
parameters, similarities between original and identified primitives,
and between corresponding combination coefficients and delays for
unconstrained data. (B) Level of variance accounted for, similarities
between original and identified primitives, and corresponding combination
coefficients and delays for non-negative data.}
\label{fig:Fig4} 
\end{figure}

Fig 5 shows the ability of then FADA and the stNMF algorithm to identify
spatiotemporal synergies and the corresponding weight coefficients
and delays from simulated non-negative mixtures, derived from model
(4) and mimicking EMG-like features. A significant main effect of
the factor Algorithm was found for the reconstruction performance
$R^{2}$ and the accuracies of the estimation of the weighting coefficients
and delays (F(4,190)≥13.34, p$<$0.001), but not for the accuracy of
the reconstruction of the primitives (F(4,190)=0.4 p$>$0.05). The factor
Noise Level had a significant main effect for $R^{2}$ and the accuracy
of the identified parameters (F(4,190)≥5.15, p$<$0.001). The interaction
of the factors Noise level and Algorithm was significant for $R^{2}$
and the accuracy of the estimation of delays (F(4,190)≥2.84, p$<$0.05).
Post-hoc testing revealed that the FADA algorithm resulted in significantly
higher $R^{2}$ values than stNMF for all noise levels (p$<$0.001).
Contrasting with this result, the FADA and stNMF algorithm provided
indistinguishable identification performance for all parameters (always
p$>$0.05), except for the identification of the delays when data were
corrupted with the highest level of noise (p=0.03). For all tested
noise levels and algorithms, $R^{2}$ and the normalized similarities
were always significantly above chance level (t(19)≥11.78,p$<$<0.001).

\begin{figure}[!h]
\begin{centering}
\includegraphics[width=0.8\columnwidth]{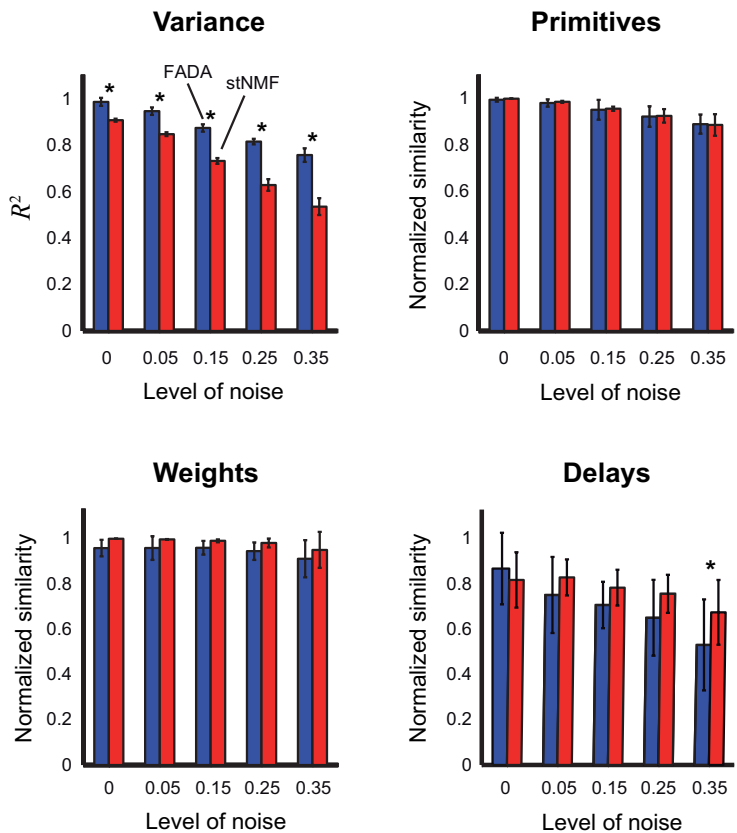} 
\par\end{centering}

\caption{Time-varying primitives. Performance of the FADA algorithm and the
identification of time-varying synergies (stNMF) for the learning
of the parameters of model (4) from ground-truth data, obtained by
combining non-negative spatiotemporal synergies. The top-left panel
shows the explained variance of the data for the two algorithms as
function of the noise level. In addition, the average similarities
between original and identified spatiotemporal primitives (top-right
panel) and the similarities of the corresponding weights and delays
(bottom-left and bottom-right panels) are shown.}
\label{fig:Fig5} 
\end{figure}

The identification performance on simulated data based on the space-by-time
generative model (5) is summarized in Fig 6. Qualitatively, the FADA
algorithm always provided better data fitting, and more precise identification
of the original temporal sources, weights and delays than the sNM3F
method. The two methods identified the spatial components with similar
levels of accuracy. In the ANOVAs the main effect of the factor Algorithm
was significant for all parameters (F(4,190)≥10.72, p$<$0.001). The
factor Noise Level had a significant main effect for $R^{2}$, as
well as on the identification accuracy of weights and delays (F(4,190)≥2.74,
p$<$0.05). A significant interaction of the two factors was found for
$R^{2}$ and for the normalized similarities associated with spatial
primitives and weighting coefficients (F(4,190)≥2.46, p$<$0.05). Post-hoc
testing showed that the FADA algorithm always provided significantly
better reconstruction of the data for all noisy data sets (p$<$0.001).
Regarding the primitives, the FADA algorithm provided more accurate
estimates of the temporal primitives for the most extreme levels of
noise (0\% and 35\%, p$<$0.01). The algorithms identified the spatial
primitives equally well (p$>$0.05). The FADA algorithm always outperformed
the sNM3F method with respect to the identification of the weighting
coefficients and temporal delays (p$<$0.05). t-tests showed that FADA
and sNM3F always provided estimates of the parameters that were better
than chance level (t(19)≥3.68, p$<$0.01), with the only exception of
the sNM3F estimation of the weighting coefficients identified from
data corrupted with 5\% of noise (t(19)=1.91, p=0.07).

\begin{figure}[!h]
\begin{centering}
\includegraphics[width=0.9\columnwidth]{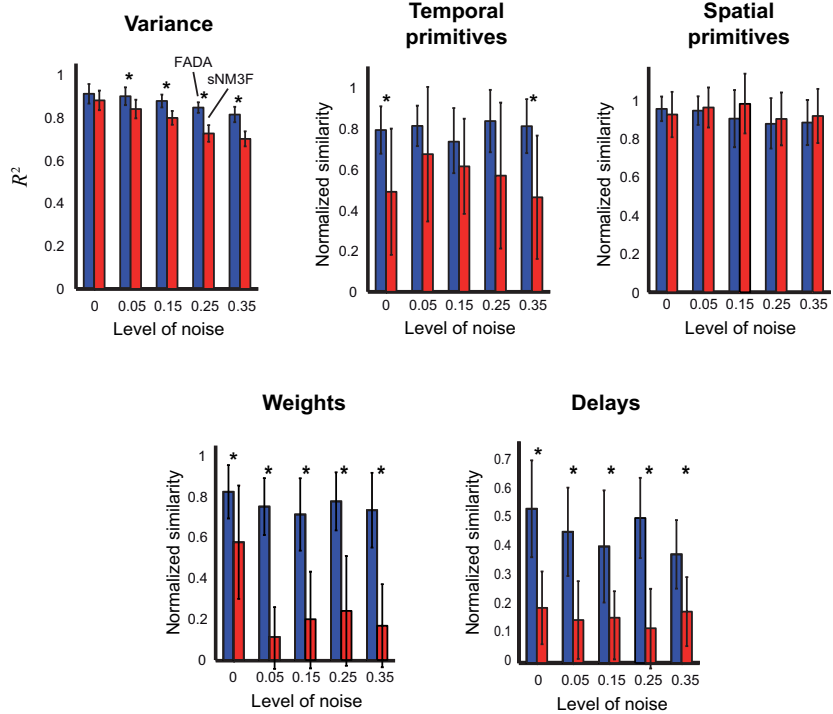} 
\par\end{centering}

\caption{Space-by-time primitives. Performance of the FADA algorithm and sampled-based
non-negative matrix tri-factorization algorithm (sNM3F) to identify
the parameters of model (5) from ground-truth data obtained combining
space-by-time synergies. The top-left graph shows the variance explained
by the two algorithms as a function of the level of noise. In addition,
the average similarities between original and identified primitives,
and the similarities of the corresponding weights and delays are shown
in the other panels. }
\label{fig:Fig6} 
\end{figure}

\subsubsection*{Comparisons of algorithm performance on real experimental data }

In addition to the validation on synthesized data, we tested the FADA
algorithm also using previously published real experimental data sets.
In addition, we compared the primitives extracted by the FADA algorithm
with those identified with other techniques. The first real experimental
data set consists of kinematic joint angle trajectories of the body
joints of participants performing emotional walks. Trajectories represented
a single gait cycle, resampled with 100 time steps {[}20, 54{]}. We
tested the FADA algorithm against the anechoic demixing algorithm
developed by Omlor and Giese {[}19{]}. Fig 7A shows that for both
algorithms the explained variance as a function of the number of primitives.
(Such plots were also used in order to determine the number of synergies,
similarly to the scree plot in statistics {[}9{]}.) The number of
primitives was identified from the $R^{2}$ curve, determining the
point where the slope levels off considerably, forming an “elbow”.
For both methods in Fig 7A this point is reached for $N$ = 3, indicating
that three anechoic components are sufficient for a reasonable approximation
of the experimental data set. Fig 7B shows also the three primitives
extracted by the two algorithms, which explain the largest amount
of variance of the data. The sources extracted with the FADA algorithm
are almost identical ($S$ = 0.96) with those extracted with the other
anechoic demixing method.

\begin{figure}[!h]
\begin{centering}
\includegraphics[width=0.55\columnwidth]{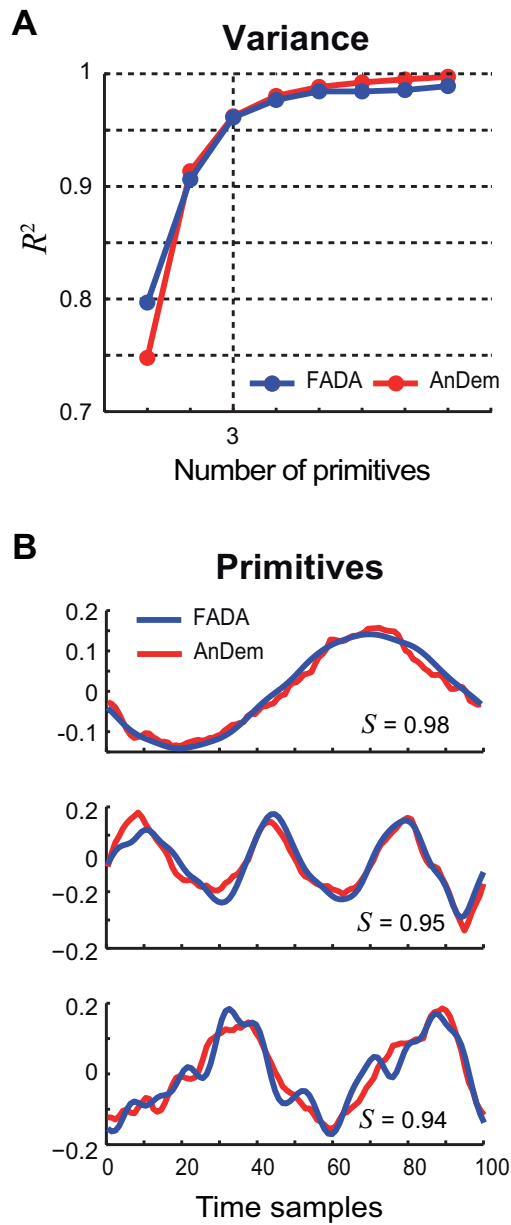} 
\par\end{centering}

\caption{Kinematic primitives extracted from experimental data. (A) Explained
variance as function of the number of extracted synergies. The blue
curve refers to the FADA algorithm and the red one to the AnDem algorithm.
(B) Temporal synergies identified by the two algorithms applied to
kinematic data collected from human participants executing emotional
walks.}
\label{fig:Fig7} 
\end{figure}

The second data set comprises EMG signals assessed during point-to-point
arm reaching movements, recording from 16 different muscles {[}9{]}.
We used the FADA and the stNMF algorithm to extract time-varying synergies.
The most-left panel in Fig 8A shows the curve obtained with the FADA
algorithm. In this case, both methods identify $N$ = 4 as levelling-off
point of the $R^{2}$ curves. The other panels in the figure show
the five time-varying synergies that were identified by the FADA algorithm.
Fig 8B shows the results obtained applying the stNMF algorithm. Similarly
to the results obtained for the FADA algorithm, the curve levels off
for $N$ = 4. The synergies of Fig 8B matched closely those in Fig
8A according to their (not normalized) level of similarity. Average
similarity across the four pairs of synergies was $S$ = 0.97 \textpm{}
0.01, indicating that, also on real EMG data, the identification performance
of the FADA algorithm was comparable to the one of the time-varying
synergies algorithm.

\begin{figure}[!h]
\begin{centering}
\includegraphics[width=0.95\columnwidth]{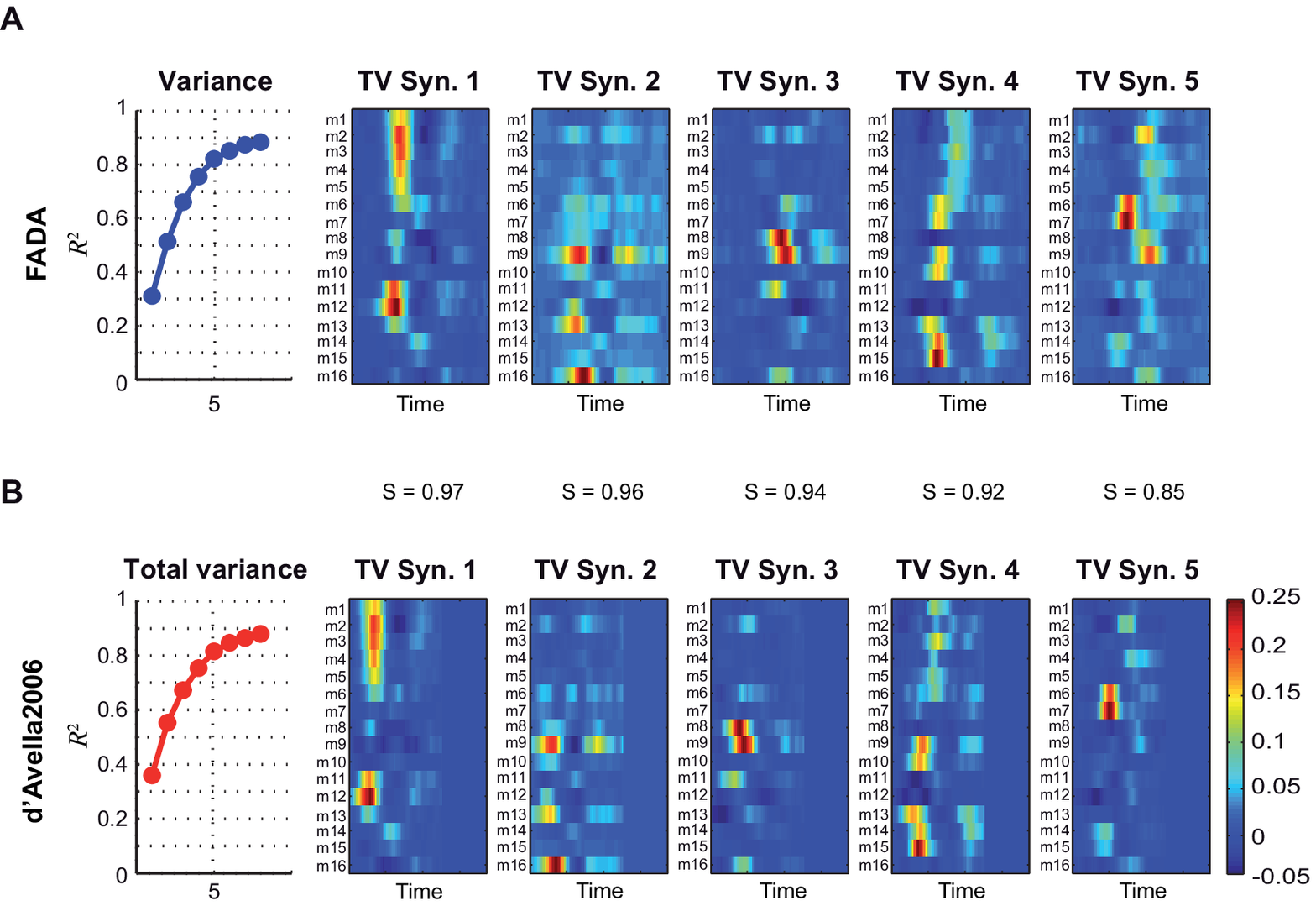} 
\par\end{centering}

\caption{EMG primitives extracted from experimental data. Spatiotemporal synergies
extracted with the FADA (A) and the stNMF algorithm (B) applied to
muscle activations collected during point-to point arm reaching movements.
Synergies are grouped according to their similarity. Most left panels
show the explained variance as function of the number of extracted
synergies for both algorithms. The dependence of this measure on the
number of extracted synergies is consistent with previously reported
data, indicating that five synergies are sufficient to account for
the significant part of the variability in the data.}
\label{fig:Fig8} 
\end{figure}

Discussion In this article, we have developed a new mathematical framework
that unifies, for the first time, many different definitions of motor
primitives. We have described how the different kinds of primitives
can be derived from a more general mixture model, which is known as
anechoic mixture, by addition of appropriate constraints. Starting
from this mathematical framework, we have implemented a new efficient
unsupervised learning algorithm for the identification of motor primitives
that achieves an identification performance typically at least as
good as the other standard methods used to study modularity in human
motor control. Such framework simplifies the comparisons between the
results from different studies using different generative models for
the definition of motor primitives. In addition, our general and robust
algorithm (Fourier Anechoic Demixing Algorithm, FADA) allows to extract
motor primitives according to specific generative models as special
cases. To promote wide adoption of the algorithm by researchers in
motor control and neurophysiology, we provide a downloadable implementation
as a MATAB toolbox. Our quantitative validation indicated that this
new algorithm performs typically equal or better than the established
methods for the extraction of primitives using different underlying
mathematical models. In the following, we discuss in detail some computational
aspects associated with the FADA algorithm and the other unsupervised
learning techniques compared with this algorithm. Moreover, taking
a broader perspective, we discuss to what extent the different definitions
of motor primitives can be really linked to a single model.

Computational considerations regarding the FADA and other unsupervised
learning algorithms As in previous studies {[}56-60{]}, we compared
FADA with other unsupervised learning techniques and assessed identification
performance on both ground-truth and experimental data sets. Differently
from all the other techniques, which are based on a single generative
model and sets of constraints for the corresponding parameters (e.g.
statistical independence or non-negativity), the FADA algorithm allows
to test different types of constraints within the same class of generative
models. In this way the proposed algorithm provides a unifying framework
for the extraction of motor primitives. A key element of the FADA
algorithm is the mapping onto a finite Fourier basis. This mapping
reduces remarkably the number of identified parameters in comparison
with more general anechoic demixing methods {[}19, 39{]}, but at the
cost that only band-limited data can be adequately modelled. For almost
all data in motor control (including at least kinematic or electromyographic
data) the informative part of the frequency spectrum typically never
exceeds 100Hz after appropriate processing. The reduction of the dimensionality
of the parameter space results in a more reliable and robust estimation
of the primitives (even in presence of substantial levels of noise)
and in a lower probability of getting stuck in local minima during
the optimization. Consequently, the FADA algorithm performed better
than other methods for the identification of anechoic primitives (Fig
4) and of temporal components associated with space-by-time decompositions
(Fig 6). The only case where it showed lower performance was the comparison
with the fastICA algorithm for the identification of unconstrained
spatial primitives using generative model (1) (Fig 2). In this case,
due to the structure of the data matrix, the data was not smooth along
the dimension that is smoothed by the FADA algorithm. In this case
the inherent smoothness prior might thus have reduced approximation
quality. In spite of this problem, the reconstruction accuracy of
the FADA algorithms was high also in this case and so were the similarity
scores for the reconstructed sources and weight matrices. For all
models including temporal delays the FADA algorithm outperformed the
other algorithms in terms of approximation quality, potentially due
the reduced number of estimated parameters. For the anechoic unconstrained
model (Figs 4), all tested algorithms achieve relatively low values
of similarities between original and identified weighting coefficients
and temporal delays, while (with the exception of Fig 6) the similarity
is significantly above chance level. Opposed to the other tested algorithms
in {[}25, 39, 48{]} we allowed for large delays and did not restrict
the tested delays to a small interval. A more detailed investigation,
which is beyond the scope of this paper, shows that the low reconstruction
accuracy is caused by ambiguities in the estimation of delays and
source functions, especially for sources with higher fundamental frequencies.
For real data the FADA algorithm provided estimates for the primitives
that were consistent with those obtained with other traditional techniques.
Also the estimated numbers of primitives for a good approximation
of the data matched between the FADA and other established algorithms
(cf. Figs 7 and 8). In addition, the functional forms of the estimated
primitives were very similar for the FADA and stNMF algorithms. Despite
its good identification performances and flexibility, the FADA algorithm
also suffers from a number of limitations. In situations where the
frequency spectrum of the real sources is not band-limited the truncated
Fourier approximation can decrease the identification performance,
as likely in the case of synchronous mixtures with few time samples
(see Fig 2). Moreover, the identification of the parameters presently
is realized by a gradient descent procedure. There exist faster optimization
methods that could be integrated in the future. For the simulations
carried out for this study the cross-correlation procedure for the
identification of the delays was implemented using entirely Matlab
built-in functions. Due to the modular architecture of FADA algorithm,
it should be easy to replace different steps by more speed-optimized
implementations and optimization methods. In this study we focused
on the design of a highly flexible rather than of a speed-optimized
algorithm.

Different definitions of motor primitives and the problem of model
selection The central mathematical contribution of this article is
that we derived how different models of motor primitives relate mathematically
to each other and how they can be derived from the anechoic mixture
model (6) by addition of appropriate constraints. This raises the
question how for a given data set the most appropriate model structure
can be found. As solutions for this model selection problem classical
criteria, such as the Akaike or the Bayesian Information criterion
(BIC) can be applied {[}61-62{]}. Alternatively, one can also use
Bayesian model selection. For this purpose, all tested models are
embedded in a joint model space, and one marginalizes the prediction
error (evidence) using an uninformative prior distribution over all
possible model architectures {[}63{]}. This procedure typically finds
automatically a good balance between the goodness-of-fit and simplicity
of the model. An implementation of this idea for automatic model selection
has been proposed in {[}63{]}, where the resulting non-Gaussian distributions
were approximated using a Laplace approximation in order to obtain
an analytically tractable selection criterion that allows to compare
different demixing models, including ones with time delays. The same
type of procedure also allows to make inferences about the most suitable
smoothness priors for a given data set.

Conclusion Inspired by the idea of Bernstein {[}64{]}, experimental
investigations in the last couple of decades have put forward the
hypothesis that the CNS might simplify the control of movement by
relying on a modular organization of control {[}1, 2{]}. The modules
(primitives) underlying such a control architecture have been defined
in multiple ways {[}65{]}, and by applying a variety of unsupervised
learning algorithms to kinematic, dynamic and EMG data sets (see for
instance {[}56{]}). This heterogeneity of approaches makes the comparison
of results across different studies very difficult. We have developed
a unifying mathematical framework for the identification of motor
primitives that links these approaches, and we have implemented a
unifying identification algorithm (FADA) that implements many different
methods as special cases as a free Matlab toolbox (FADA) that is available
online. We demonstrated that the FADA algorithm typically shows identification
performance that is competitive with other classical unsupervised
learning techniques. In some cases, it even outperforms these techniques
for data from motor control, especially in presence of noise. We hope
that the new Matlab toolbox will help to establish more solid links
between different definitions of motor primitives, helping neuroscientists
with the comparison between different theoretical models and their
data.

\section*{Acknowledgments}

We thank Lars Omlor for his help during the first stage of this project,
and Dominik Endres for many helpful discussions. The research leading to these results has received funding
from the European Union Horizon 2020 Programme
(H2020/2014-2020) under grant agreement n H2020 ICT-23-2014 /644727 
Cogimon (www.cogimon.eu).
Martin Giese and Enrico Chiovetto have been also supported by the following
funding sources: FP7-ICT-2013-10 (Koroibot);
DFG GI 305/4-1, DFG GZ: KA 1258/15-1; BMBF, FKZ:
01GQ1002A, FP7-PEOPLE-2011-ITN (Marie Curie): ABC
PITN-GA-011-290011; FP7/2007-2013/604102 (HBP).

\section*{Bibliography}

1. Bizzi E, Cheung VCK, d'Avella A, Saltiel P, Tresch M. Combining
modules for movement. Brain Research Reviews. 2008 Jan; 57(1): p.
125-133.\\
 2. Flash T, Hochner B. Motor primitives in vertebrates and invertebrates.
Curr Opin Neurobiol. 2005 Dec; 15(6): p. 660-666.\\
 3. Tresch MC, Saltiel P, Bizzi E. The construction of movement by
the spinal cord. Nat Neurosci. 1999 Feb; 2(2): p. 162-167.\\
 4. Berret B, Bonnetblanc F, Papaxanthis C, Pozzo T. Modular control
of pointing beyond arm's length. J Neurosci. 2009 Jan; 29(1): p. 191-205.\\
 5. Kaminski TR. The coupling between upper and lower extremity synergies
during whole body reaching. Gait Posture. 2007 Jul; 26(2): p. 256-262.\\
 6. Thomas JS, Corcos DM, Hasan Z. Kinematic and kinetic constraints
on arm, trunk, and leg segments in target-reaching movements. J Neurophysiol.
2005 Jan; 93(1): p. 352-364.\\
 7. Mussa-Ivaldi FA, Giszter SF. Vector field approximation: a computational
paradigm for motor control and learning. Biol Cybern. 1992; 67(6):
p. 491-500.\\
 8. Chiovetto E, Berret B, Pozzo T. Tri-dimensional and triphasic
muscle organization of whole-body pointing movements. Neuroscience.
2010 Nov; 170(4): p. 1223-1238.\\
 9. d'Avella A, Portone A, Fernandez L, Lacquaniti F. Control of fast-reaching
movements by muscle synergy combinations. J Neurosci. 2006 Jul; 26(30):
p. 7791-7810.\\
 10. Torres-Oviedo G, Macpherson JM, Ting LH. Muscle synergy organization
is robust across a variety of postural perturbations. J Neurophysiol.
2006 Sep; 96(3): p. 1530-1546.\\
 11. Ivanenko YP, Poppele RE, Lacquaniti F. Five basic muscle activation
patterns account for muscle activity during human locomotion. J Physiol.
2004 Apr; 556(Pt 1): p. 267-282.\\
 12. Ivanenko YP, Cappellini G, Dominici N, Poppele RE, Lacquaniti
F. Coordination of locomotion with voluntary movements in humans.
J Neurosci. 2005 Aug; 25(31): p. 7238-7253.\\
 13. Chiovetto E, Patanè L, Pozzo T. Variant and invariant features
characterizing natural and reverse whole-body pointing movements.
Exp Brain Res. 2012 May; 218(3): p. 419-431.\\
 14. Santello M, Flanders M, Soechting JF. Postural hand synergies
for tool use. J Neurosci. 1998 Dec; 18(23): p. 10105-10115.\\
 15. d'Avella A, Saltiel P, Bizzi E. Combinations of muscle synergies
in the construction of a natural motor behavior. Nat Neurosci. 2003
Mar; 6(3): p. 300-308.\\
 16. Torres-Oviedo G, Ting LH. Muscle synergies characterizing human
postural responses. J Neurophysiol. 2007 Oct; 98(4): p. 2144-2156.\\
 17. Cheung VCK, d'Avella A, Tresch MC, Bizzi E. Central and sensory
contributions to the activation and organization of muscle synergies
during natural motor behaviors. J Neurosci. 2005 Jul; 25(27): p. 6419-6434.\\
 18. Chiovetto E, Giese MA. Kinematics of the Coordination of Pointing
during Locomotion. PLoS One. 2013; 8(11): p. e79555.\\
 19. Omlor L, Giese MA. Anechoic blind source separation using Wigner
marginals. Journal of Machine Learning Research. 2011 Mar; 12: p.
1111-1148.\\
 20. Roether CL, Omlor L, Christensen A, Giese MA. Critical features
for the perception of emotion from gait. J Vis. 2009; 9(6): p. 15.1-1532.\\
 21. Omlor L, Giese MA. Blind source separation for over-determined
delayed mixtures. In: B. Schölkopf, J. Platt, and T. Hoffman, editors.
Advances in Neural Information Processing Systems 19. Cambridge, MA:
MIT Press; 2007. p. 1049-1056.\\
 22. Ting LH, Macpherson JM. A limited set of muscle synergies for
force control during a postural task. J Neurophysiol. 2005 Jan; 93(1):
p. 609-613.\\
 23. Omlor L, Giese MA. Extraction of spatio-temporal primitives of
emotional body expressions. Neurocomputing, 2007; 70(10-12), p. 1938-1942.
\\
 24. d'Avella A, Tresch MC. Modularity in the motor system: decomposition
of muscle patterns as combinations of time-varying synergies. In:
Solla SA, editor. Advances in Neural Information Processing Systems
14. Cambridge, MA: MIT Press; 2002. p. 141-148.\\
 25. Delis I, Panzeri S, Pozzo T, Berret B. A unifying model of concurrent
spatial and temporal modularity in muscle activity. J Neurophysiol.
111(3), p. 675-693.\\
 26. Bofill P. Underdetermined blind separation of delayed sound sources
in the frequency domain. Neurocomputing. 2003; 55(3): p. 627-641.\\
 27. Emile B, Comon P. Estimation of time delays between unknown colored
signals. Signal Processing. 1998; 69(1): p. 93-100.\\
 28. Torkkola K. Blind separation of delayed sources based on information
maximization. In: IEEE International Conference on Acoustics, Speech,
and Signal Processing. 1996; 6, pp. 3509-3512. \\
 29. Yilmaz O, Rickard S. Blind separation of speech mixtures via
time-frequency masking. IEEE transactions on Signal Processing. 2004
Jul;52(7):1830-47. \\
 30. Choi S, Cichocki A, Park HM, Lee SY. Blind Source Separation
and Independent Component Analysis: A Review. Neural Information Processing
- Letters and Reviews. 2005; 6(1): p. 1-57.\\
 31. Comon P, Jutten C, editors. Handbook of Blind Source Separation:
Independent component analysis and applications. Academic press; 2010.\\
 32. Bell AJ, Sejnowski TJ. An information-maximization approach to
blind separation and blind deconvolution. Neural Comput. 1995 Nov;
7(6): p. 1129-1159.\\
 33. Be'ery E, Yeredor A. Blind separation of superimposed shifted
images using parameterized joint diagonalization. IEEE Trans Image
Process. 2008 Mar; 17(3): p. 340-353.\\
 34. Lee DD, Seung HS. Algorithms for non-negative matrix factorization.
In Advances in neural information processing systems. 2001: p. 556-562.\\
 35. O'Grady PD, Pearlmutter BA, Rickard ST. Survey of sparse and
non-sparse methods in source separation. International Journal of
Imaging Systems and Technology. 2005; 15(1): p. 18-33.\\
 36. Arberet S, Gribonval R, Bimbot F. A robust method to count and
locate audio sources in a stereophonic linear instantaneous mixture.
In Independent Component Analysis and Blind Signal Separation. Springer;
2006. p. 536-543.\\
 37. Cho N, Kuo CC. Underdetermined audio source separation from anechoic
mixtures with long time delay. In Acoustics, Speech and Signal Processing,
2009. ICASSP 2009. IEEE International Conference on 2009 Apr 19 p.
1557-1560.\\
 38. Harshman RA, Hong S, Lundy ME. Shifted factor analysis—Part I:
Models and properties. Journal of chemometrics. 2003; 17(7): p. 363-378.\\
 39. Mørup M, Madsen KH, Hansen LK. Shifted independent component
analysis. In: Independent Component Analysis and Signal Separation
2007 Jan 1 (pp. 89-96). Springer Berlin Heidelberg.\\
 40. Højen-Sørensen PA, Winther O, Hansen LK. Mean-field approaches
to independent component analysis. Neural Computation. 2002 Apr;14(4):889-918.\\
 41. Lee DD, Seung HS. Learning the parts of objects by non-negative
matrix factorization. Nature. 1999 Oct; 401(6755): p. 788-791.\\
 42. Swindlehurst A. Time Delay and Spatial Signature Estimation Using
Known Asynchronous Signals. IEEE Trans. Signal Processing. 1997; 46:
p. 449-462.\\
 43. d'Avella A, Bizzi E. Shared and specific muscle synergies in
natural motor behaviors. Proceedings of the National Academy of Sciences
of the United States of America. 2005; 102(8): p. 3076-3081.\\
 44. Mallat SG, Zhang Z. Matching pursuits with time-frequency dictionaries.
Signal Processing, IEEE Transactions on. 1993; 41(12): p. 3397-3415.\\
 45. Hyvarinen A. Fast and robust fixed-point algorithms for independent
component analysis. IEEE Transactions on Neural Networks. 1999 May;
10(3): p. 626-634.\\
 46. Hyvärinen A, Oja E. A Fast Fixed-Point Algorithm for Independent
Component Analysis. Neural Computation. 1997; 9(7): p. 1483-1492.\\
 47. M{ø}rup M, Madsen KH. SICA, 2007. 48. M{ø}rup M, Madsen KH,
Hansen LK. Shifted non-negative matrix factorization. In: IEEE Workshop
on Machine Learning for Signal Processing. 2007; pp. 139-144.\\
 49. Kass RE, Ventura V. A Spike-Train Probability Model. Neural Computation.
2001; 13(8): p. 1713-1720.\\
 50. Harris CM, Wolpert DM. Signal-dependent noise determines motor
planning. Nature. 1998 Aug; 394(6695): p. 780-784.\\
 51. Schmidt RA, Zelaznik H, Hawkins B, Frank JS, Quinn JT. Motor-output
variability: a theory for the accuracy of rapid motor acts. Psychol
Rev. 1979 Sep; 47(5): p. 415-451.\\
 52. Sutton GG, Sykes K. The variation of hand tremor with force in
healthy subjects. J Physiol. 1967 Aug; 191(3): p. 699-711.\\
 53. van RJ, Haggard P, Wolpert DM. The role of execution noise in
movement variability. J Neurophysiol. 2004 Feb; 91(2): p. 1050-1063.\\
 54. Endres D, Chiovetto E, Giese MA. Model selection for the extraction
of movement primitives. Frontiers in Computational Neuroscience. 2013;
7(185).\\
 55. Mardia KV, Kent JT, Bibby JM. Multivariate Analysis: Academic
Press; 1979.\\
 56. Tresch MC, Cheung VCK, d'Avella A. Matrix factorization algorithms
for the identification of muscle synergies: evaluation on simulated
and experimental data sets. J Neurophysiol. 2006; 95(4): p. 2199-2212.\\
 57. Cashero Z, Anderson C. Comparison of EEG blind source separation
techniques to improve the classification of P300 trials. Conf Proc
IEEE Eng Med Biol Soc. 2011; 2011: p. 7183-7186.\\
 58. Caulo M, Esposito R, Mantini D, Briganti C, Sestieri C, Mattei
PA, et al. Comparison of hypothesis- and a novel hybrid data/hypothesis-driven
method of functional MR imaging analysis in patients with brain gliomas.
AJNR Am J Neuroradiol. 2011; 32(6): p. 1056-1064.\\
 59. Erhardt EB, Rachakonda S, Bedrick EJ, Allen EA, Adali T, Calhoun
VD. Comparison of multi-subject ICA methods for analysis of fMRI data.
Hum Brain Mapp. 2011 Dec; 32(12): p. 2075-2095.\\
 60. Virtanen J, Noponen T, Meriläinen P. Comparison of principal
and independent component analysis in removing extracerebral interference
from near-infrared spectroscopy signals. J Biomed Opt. 2009; 14(5):
p. 054032.\\
 61. Akaike H. A new look at the statistical model identification.
Automatic Control, IEEE Transactions on. 1974 dec; 19(6): p. 716-723.
62. Schwarz G. Estimating the dimension of a model. The Annals of
Statistics. 1978; 6: p. 461-464.\\
 63. Bishop CM. Pattern Recognition and Machine Learning: Springer;
2007.\\
 64. Bernstein N. The coordination and regulation of movements.: Oxford:
Pergamon; 1967.\\
 65. Chiovetto E, Berret B, Delis I, Panzeri S, Pozzo T. Investigating
reduction of dimensionality during single-joint elbow movements: a
case study on muscle synergies. Front Comput Neurosci. 2013; 7: p.
11.\\

\end{document}